\documentclass[prb,aps,epsf,showpacs,twocolumn,longbibliography]{revtex4-1}
\usepackage{times}
\usepackage{graphicx}
\usepackage{float}
\usepackage{latexsym,amsmath,amssymb,bm,euscript}
\usepackage{color}
\usepackage{subfigure}
\usepackage{epstopdf}
\usepackage[colorlinks=true,linkcolor=blue,citecolor=blue,urlcolor=blue]{hyperref}
\usepackage{hyperref}
\usepackage{ulem}
\usepackage[dvipsnames]{xcolor}

\begin{document}

\title{Density Matrix Renormalization Group Study of Nematicity in Two Dimensions: \\ Application to a Spin-$1$ Bilinear-Biquadratic Model on the Square Lattice}
\author{Wen-Jun Hu$^{1}$}
\author{Shou-Shu Gong$^2$}
\email{shoushu.gong@buaa.edu.cn}
\author{Hsin-Hua Lai$^{3}$}
\author{Qimiao Si$^{3}$}
\email{qmsi@rice.edu}
\author{Elbio Dagotto$^{1,4}$}
\email{edagotto@utk.edu}
\affiliation{
$^1$ Department of Physics and Astronomy, University of Tennessee, Knoxville, Tennessee 37996, USA\\
$^2$ Department of Physics, Beihang University, Beijing 100191, China\\
$^3$ Department of Physics and Astronomy \& Rice Center for Quantum Materials, Rice University, Houston, Texas 77005, USA\\
$^4$ Materials Science and Technology Division, Oak Ridge National Laboratory, Oak Ridge, Tennessee 37831, USA
}

\begin{abstract}
Nematic order is an exotic property observed in several strongly correlated systems, such as the iron-based superconductors.
Using large-scale density matrix renormalization group (DMRG) techniques, we study at zero-temperature the
nematic spin liquid that competes with spin dipolar and quadrupolar orders.  
We use these nematic orders to characterize different quantum phases and quantum phase transitions.
More specifically, we study a spin-$1$ bilinear-biquadratic Heisenberg model on the square lattice with couplings beyond nearest neighbors. 
We focus on parameter regions around the highly symmetric $SU(3)$ point where the bilinear and biquadratic interactions are equal.
With growing further-neighbor biquadratic interactions, we identify different spin dipolar and quadrupolar orders.
We find that the DMRG results on cylindrical geometries correctly detect nematicity in different quantum states and accurately characterize the quantum phase transitions among them.
Therefore, spin-driven nematicity -- here defined as the spontaneous breaking of the lattice invariance under a 90$^o$ rotation -- 
is an order parameter which can be studied directly in DMRG calculations in two 
dimensions in different quantum states.
\end{abstract}


\maketitle

\section{Introduction}

Rotational invariance is one of the fundamental lattice symmetries in condensed matter crystals.
The spontaneous breaking of this rotational symmetry can occur via the emergence of magnetic nematicity, 
which may lead to novel phases. Nematic phases also occur in liquid crystals.
In recent years, nematicity has also been found in strongly correlated electronic systems, 
including the fractional quantum Hall effect~\cite{musaelian1996,balents1996,Mulligan2010,Mulligan2011,Maciejko2013,Liu2013,You2014,Gromov2017,du2019} and the iron-based superconductors~\cite{Kivelson2008,Sachdev2008,bao2009,Shiliang2009,Dai4118,Chu824,dai2012magnetism,liang2013,fernandes2014drives,Lu657,Bishop2016,Bishop2017}
In such systems, nematicity is usually observed accompanied with novel quantum states or new collective excitations, such as the nematic phase without long-range spin order in the iron-based superconductor~\cite{Bishop2016} and the emergent new plasmon excitations in the fractional quantum Hall state~\cite{du2019}. 
Therefore, studying the origin of nematicity and its interplay with other properties in strongly correlated systems is important not only for understanding the novel states but also for predicting new quantum phases.

Very recently, nematicity was also found at zero temperature 
in a quantum spin liquid phase with spontaneous lattice $C_4$ rotational symmetry breaking~\cite{hunsl}.  
This nematic spin liquid was identified in a spin-$1$ model on the square lattice with bilinear-biquadratic interactions, which is defined as
\begin{equation}\label{model}
H = \sum_{i,j} J_{ij} {\bf S}_i \cdot {\bf S}_j + K_{ij} ( {\bf S}_i \cdot {\bf S}_j )^2,
\end{equation}
where ${\bf S}_{i}$ is a spin-$1$ operator at site $i$, while $J_{ij}$ and $K_{ij}$ are the bilinear  and biquadratic interactions, respectively~\cite{papanicolaou1988}. 
Similar bilinear-biquadratic models were studied to describe spin-$1$ magnetic systems, such as the triangular-lattice layered materials NiGa$_2$S$_4$~\cite{nakatsuji2005} and Ba$_3$NiSb$_2$O$_9$~\cite{cheng2011, fak2017}, the honeycomb-lattice 6HB-Ba$_3$NiSb$_2$O$_9$~\cite{quilliam2016}, 
and the iron-pnictide~\cite{rong2012} and iron-chalcogenide superconductors~\cite{rong2015,wang2016,gong2017,lai2017}.
Thus, such spin-$1$ bilinear-biquadratic Hamiltonian is a prototypical model to search for novel quantum phases in frustrated spin-$1$ systems~\cite{Haldane1983_2, aklt1987, katsumata1989, hagiwara1990, white1993, schollwock1996, shelton1996, lauchli2006, corboz2007, corboz2017}.

In one dimension, the ground state phase diagram of Eq.~\eqref{model} with nearest-neighbor couplings has been well established~\cite{lauchli2006}.
A symmetry-protected topological phase has been identified, known as the Haldane state, with a finite bulk gap and gapless edge states if open boundaries are used~\cite{Haldane1983_1, Haldane1983_2}. 
In two dimensions (2D), according to the Lieb-Schultz-Mattis-Hastings theorem~\cite{lieb1961,Hastings2004}, the ground state of such a spin-$1$ model could be either a conventional ordered state with spontaneous symmetry breaking, a gapless spin liquid, a gapped spin liquid with topological order, or a quantum paramagnet with a unique ground state and finite bulk gap. 

With these many fruitful possibilities, the studies of novel quantum phases in such 2D systems have attracted much attention, especially near the highly symmetric $SU(3)$ point with equal bilinear and biquadratic interactions $J_{ij} = K_{ij}$~\cite{changlani,lauchli2006_2,toth2010, toth2012,bauer2012,zhao2012,corboz2012,corboz2013,corboz2017,niesen2017tensor,corboz2018}, which has enhanced frustration due to the strong competition between the two types of interactions.
By only considering the nearest-neighbor couplings, a three-sublattice magnetic order has been reported near the $SU(3)$ point on the triangular~\cite{lauchli2006_2,bauer2012,corboz2018} and square lattices~\cite{toth2010, toth2012,bauer2012,corboz2017,niesen2017tensor}.
Alternatively, the valence bond solid states, which spontaneously break lattice symmetries including a trimerized ground state on the kagome lattice~\cite{changlani,corboz2012} and a plaquette state on the honeycomb lattice~\cite{zhao2012,corboz2013}, have also been suggested to exist near the $SU(3)$ point. 
Most surprisingly, recent large-scale density matrix renormalization group (DMRG) calculations confirmed the three-sublattice order in the triangular $SU(3)$ model, but found new evidence to support a nematic quantum spin liquid rather than the three-sublattice magnetic order in the square $SU(3)$ model~\cite{hunsl}.
This nematic spin liquid is particularly interesting since it does not have either spin dipolar or quadrupolar long-range order, but has a nonzero nematic order with $C_4$ lattice rotational symmetry broken spontaneously while it preserves translational symmetry~\cite{hunsl}.
 
Nematicity has rarely been reported in quantum spin liquid states, and its driving force in this setting is still to be understood.
In Ref.~\onlinecite{hunsl}, the nematicity in the spin liquid phase was argued to be caused by 
the fluctuations peaked at the momentum $(\pi, 2\pi/3)$ on a finite-size cluster.
These dominant fluctuations appear clearly in the spin and quadrupolar static structure factors, and are proposed to be the result of melting order at momentum $(\pi, 2\pi/3)$. Note that the state with wavevector $(2\pi/3,\pi)$ in a two-dimensional 
system is totally equivalent: starting from high temperature and cooling down via an annealing or Monte Carlo process, if it
were possible, both states have equal chance, similarly as the $(\pi,0)$ and $(0,\pi)$ states in undoped iron superconductors. The
cylindrical boundary conditions used here may introduce a small symmetry breaking preference for $(\pi,2\pi/3)$, like a small magnetic field up favors in a ferromagnetic Ising model the state with all spins up. But these are minor details that do not affect our nematic analysis: there are two wavevector states quasi-degenerate and only one can become the ground state.

The tendency towards $(\pi,2\pi/3)$ fluctuations can also be discussed using a coupled-chain picture of a conjectured gapless spin liquid; here $2\pi/3$ is twice the parton ``Fermi-wavevector'' for an $SU(3)$ chain, which is associated with the $1/3$-filling of those partons, and $\pi$ simply 
reflects the interchain coupling that is relevant in the RG sense~\cite{hunsl}.
However, the precise origin of the nematicity and dominance of the wavevectors $(\pi,2\pi/3)$ or $(2\pi/3,\pi)$ still requires further investigations.
Before answering these difficult questions, it is important to establish how nematicity 
is stabilized in different quantum phases and how does it behave at quantum phase transitions.

In this paper, we use large-scale DMRG calculations to address these issues by simulating the model ~\eqref{model} while adding further-neighbor biquadratic interactions.
We focus on parameter regions around the $SU(3)$ point with antiferromagnetic $J_1 = K_1 = 1.0$ ($J_1$ and $K_1$ are the nearest-neighbor bilinear and biquadratic interactions, respectively). 
We show that the small nematic spin liquid region reported recently near the $SU(3)$ point with $K_1 = J_1$~\cite{hunsl} can actually be broadened by adding the second-neighbor ($K_2$) and third-neighbor ($K_3$) biquadratic interactions.
In addition, with further growing biquadratic interaction strengths, different spin dipolar and quadruplar orders are identified, characterized by specific spin and quadrupolar correlation functions and structure factors.
For better describing the nematicity, we study both the $B_{1g}$ and $B_{2g}$ nematic order channels that have been proposed to classify different rotational symmetry breakings on the square lattice~\cite{b1gb2g}. 

Through our extensive DMRG simulations, a plethora of new quantum phases are unveiled here, including quantum phases with either $B_{1g}$ or $B_{2g}$ nematicity, as well as phases that preserve lattice rotational symmetry. 
We find that the obtained nematic orders faithfully describe the quantum states and the corresponding quantum phase transitions.
Interestingly, although our DMRG simulations are performed on cylindrical geometries, which {\it a priori} explicitly break lattice rotational invariance, we show that this explicit breaking leads to small effects for the systems we study. On the other hand, when nematic order parameters are studied in regions where nematicity appears stable the signal is very robust, much larger than the small effect caused by the geometry. As a consequence, DMRG studies using cylinders can be used to identify quantum phase transitions involving nematicity, in addition to the standard analysis of the spin-spin and quadrupolar-quadrupolar correlation functions. Our work thus provides a procedure to use the DMRG to analyze nematic orders and nematic phase transitions in spin square-lattice models, which can be generalized to other geometries.

Our paper is organized as follow.
In Sec.~\ref{sec:method}, we introduce computational details and define the order parameters that we will measure.
In Sec.~\ref{sec:k2} and Sec.~\ref{sec:k3}, we show the quantum phase diagrams of model Eq.~\eqref{model} varying $K_2$ or $K_3$, and we characterize different spin dipolar and quadrupolar phases by their correlation functions and order parameters.
In Sec.~\ref{sec:nematic}, we calculate nematic orders for different phases and use these nematic orders to characterize the quantum phase transitions.
A summary and discussion is given in Sec.~\ref{sec:sum}.

\section{Method and Order Parameters}
\label{sec:method}

\begin{figure}[t]
\includegraphics[width = 1\linewidth]{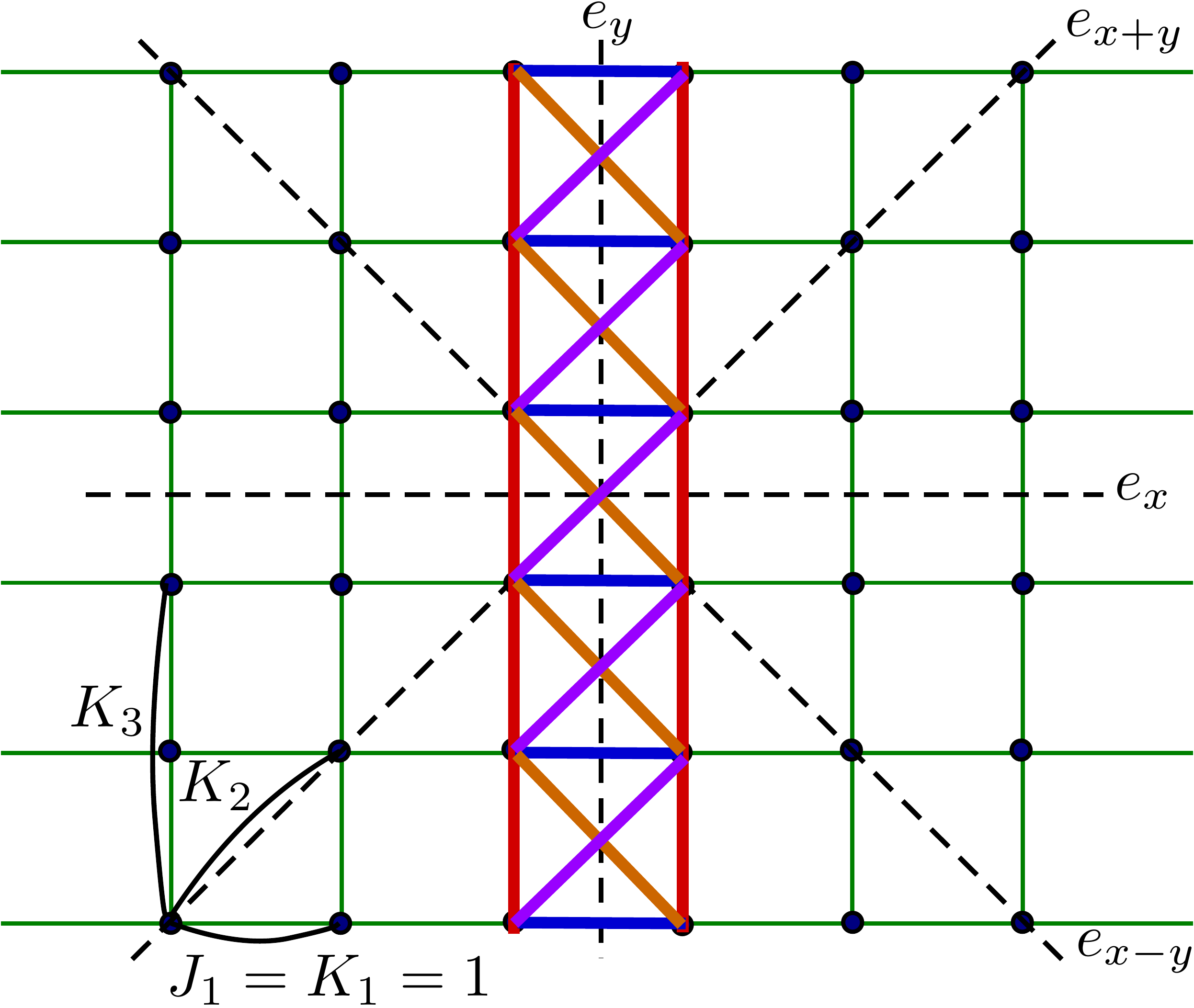}
\caption{Sketch illustrating bonds in the middle of the RC6 cylinder used for calculating the nematic order parameters in both the $B_{1g}$ and $B_{2g}$ channels. The dashed lines represent mirror symmetries along $e_x$, $e_{y}$, $e_{x+y}$, and $e_{x-y}$. The blue, red, purple, and orange lines represent bonds along the $e_x$, $e_{y}$, $e_{x+y}$, and $e_{x-y}$ directions, respectively.}
\label{nematicity}
\end{figure}

In the $SU(2)$ DMRG simulation~\cite{white1992, mcculloch2002}, we set the NN bilinear coupling $J_1 = 1.0$ as the energy scale and we focus on the parameter regions around the $SU(3)$ point with $K_1 = J_1$.
We will consider additional second-neighbor ($K_2$) or third-neighbor ($K_3$) biquadratic interactions (see Fig.~\ref{nematicity}) to study the quantum phases of the system Eq.~\eqref{model} on the square lattice. By adding extra couplings we can enhance the 
region of stability of the spin liquid found at the SU(3) point, rendering its properties more clear.
To reduce the influence of finite-size effects as much as possible, we study the system on both the rectangular cylinder (RC) and the $45^\circ$-tilted cylindrical (TC) geometries.
These two cylindrical geometries have periodic boundary conditions in the $y$ direction (circumference direction) and open boundaries in the $x$ direction (axis direction), and we denote them as RC$L_y$ or TC$L_y$ ($L_y$ is the number of sites along 
the short circumference direction). With regards to the ``long'' direction along the axis, we analyzed systems with $L_x=24$ and $36$ to confirm that we obtain similar results.
Most of our DMRG simulations have focused on the RC6 geometry, which can harbor both three- and two-sublattice order structures. 
Also, we selected some typical points to perform DMRG simulations on RC9 clusters (because they
are highly demanding of computer time) to check the stability of the several quantum phases on larger clusters.
For three-sublattice orders, we also study the TC geometry for cross-checking purposes.
By keeping up to $4000$ $SU(2)$ DMRG states, most of the truncation errors in the important intermediate coupling region are about
$10^{-7}$, while the largest ones in our calculations are around $10^{-5}$.

To characterize different quantum states, we calculate spin $\langle {\bf S}_{i} \cdot {\bf S}_{j} \rangle $ and quadrupolar $ \langle {\bf Q}_{i} \cdot {\bf Q}_{j} \rangle$ correlation functions as well as their structure factors, where the quadrupolar tensor operator is ${\bf Q}_{i}=\left( \tfrac{1}{2}(Q_i^{xx}-Q_i^{yy}), \tfrac{1}{2\sqrt{3}}(2Q_i^{zz}-Q_i^{xx}-Q_i^{yy} ), Q_i^{xy}, Q_i^{yz}, Q_i^{xz}\right)$~\cite{blume1969,toth2010}, with $Q_i^{\alpha \beta}=S_i^\alpha S_i^\beta + S_i^\beta S_i^\alpha-\tfrac{4}{3} \delta_{\alpha \beta}$ \mbox{($\alpha,\beta\!=\!x,y,z$)}. The quadrupolar correlation function can be expressed as ${\bf Q}_{i}\cdot {\bf Q}_{j}=2({\bf S}_{i}\cdot {\bf S}_{j})^2+{\bf S}_{i}\cdot {\bf S}_{j}-8/3$. We perform the Fourier transformation for the correlation functions to obtain the spin structure factor as 
\begin{equation}
m^{2}_{S}({\bf q}) = \frac{1}{N_s^2}\sum_{i,j} \langle {\bf S}_{i}\cdot {\bf S}_{j} \rangle e^{i{\bf q}\cdot({\bf r}_i-{\bf r}_j)},
\end{equation}
and the quadrupolar structure factor as
\begin{equation}
m^{2}_{Q}({\bf q}) = \frac{1}{N_s^2}\sum_{i,j} \langle {\bf Q}_{i}\cdot {\bf Q}_{j} \rangle e^{i{\bf q}\cdot({\bf r}_i - {\bf r}_j)},
\end{equation}
where the sites $i, j$ are chosen inside the middle region of size $N_s = L_y \times 2L_y$ in order to avoid the effects of open edges and also be able to consider both two- and three-sublattice orders~\cite{hunsl}.
We use these correlation functions and structure factors to detect spin dipolar and quadrupolar orders.
As the range of accessible system sizes available for DMRG spin-$1$ systems is limited due to the difficulty
of the calculation, we cannot perform detailed finite-size scaling analysis.
Fortunately, in our studied parameter regions the measured spin dipolar and quadrupolar orders are strong, 
and for this reason they can be clearly identified on the $L_y = 6, 9$ finite systems we employ. 

\begin{figure}[t]
\includegraphics[width = \linewidth]{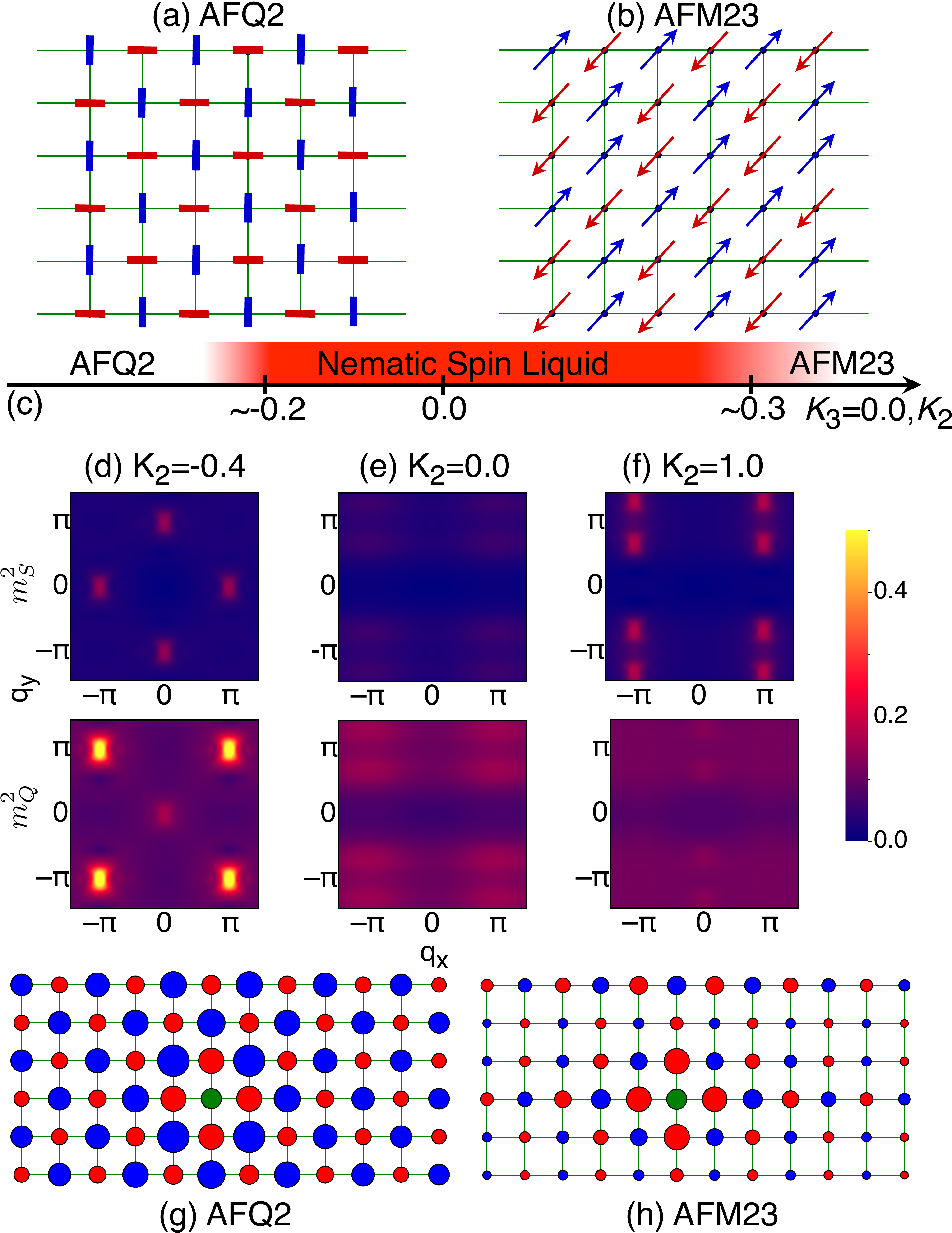}
\caption{Quantum phase diagram of the $SU(3)$ model with additional $K_2$ interaction, at $K_3 = 0$.
(a) and (b) are sketches illustrating the AFQ2 and AFM23 orders. 
(c) Quantum phase diagram of the spin-1 $SU(3)$ Heisenberg model on the square lattice 
with $J_1=K_1=1.0, K_3 = 0$, and varying $K_2$. 
The regime of the nematic spin liquid around the $SU(3)$ point is indicated by red shading. 
(d)-(f) are spin ($m^2_S$) and quadrupolar ($m^2_Q$) structure factors for different values of $K_2$, which are obtained from the middle $6\times 12$ sites region of a long RC6 cylinder. 
The upper and lower figures are for $m^2_S$ and $m^2_Q$, respectively. 
(g) The real space quadrupolar correlation functions for the AFQ2 state with $K_2=-0.4$ in the middle of the RC6 cylinder. (h) The real space spin correlation functions for the AFM23 state with $K_2=1.0$ in the middle of the RC6 cylinder. 
The green site is the reference site; the blue and red colors denote positive and negative correlations of the sites with respect to the reference site, respectively. The areas of circles are proportional to the magnitude of the correlations.}\label{pdk2}
\end{figure}

\begin{figure}[t]
\includegraphics[width = \linewidth]{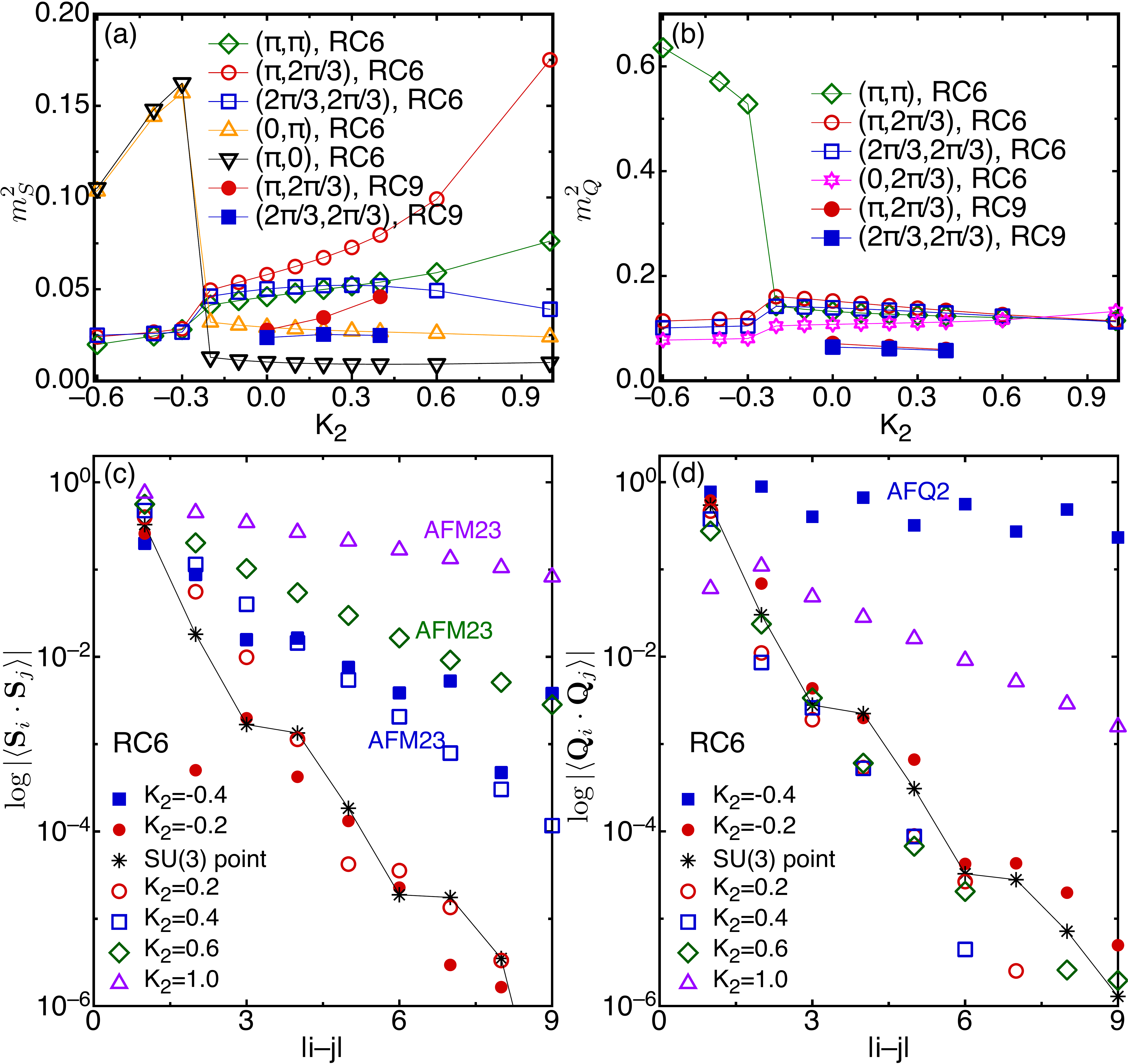}
\caption{Interaction dependence of order parameters and correlation functions 
in the $SU(3)$ model with additional $K_2$ coupling.
(a) and (b) are spin ($m^2_S$) and quadrupolar ($m^2_Q$) order parameters 
for $J_1=K_1=1.0, K_3=0.0$, at different momenta as a function of $K_2$ on the RC6 and RC9 cylinders. 
Both order parameters are obtained from the middle $6\times 12$ sites on 
the RC6 cylinder or the middle $9\times 18$ sites on the RC9 cylinder. 
(c) and (d) are the log-linear plots of spin $\langle {\bf S}_i\cdot{\bf S}_j \rangle$ 
and quadrupolar $\langle {\bf Q}_i\cdot{\bf Q}_j \rangle$ correlations as a function of the distance on the RC6 cylinder.}
\label{j1k1k2}
\end{figure}

In some of the magnetic order phases we investigated, the lattice rotational symmetry by 90$^o$ is also broken (nematicity) accompanied by spin rotational symmetry breaking. 
To study better this nematicity, we define the lattice nematic order parameters in both the $B_{1g}$ and $B_{2g}$ channels for both spin dipolar and quadrupolar operators~\cite{Chandra1990,rong2015,hunsl,b1gb2g}, as shown in Fig.~\ref{nematicity}.
Therefore, we have four nematic order parameters with two of them in the $B_{1g}$ channel for the NN bonds as
\begin{eqnarray}
\sigma^S_{B1g} &=& \frac{1}{N_m}\sum_i[\langle {\bf S}_{i}\cdot {\bf S}_{i+\hat{x}}\rangle - \langle {\bf S}_{i}\cdot {\bf S}_{i+\hat{y}}\rangle] \label{sigma1s}, \\
\sigma^Q_{B1g} &=& \frac{1}{N_m}\sum_i[\langle {\bf Q}_{i}\cdot {\bf Q}_{i+\hat{x}}\rangle - \langle {\bf Q}_{i}\cdot {\bf Q}_{i+\hat{y}}\rangle] \label{sigma1q},
\end{eqnarray}
and the other two parameters in the $B_{2g}$ channel for the NNN bonds as
\begin{eqnarray}
\sigma^S_{B2g} &=& \frac{1}{N_m}\sum_i[\langle {\bf S}_{i}\cdot {\bf S}_{i+\hat{x}+\hat{y}}\rangle - \langle {\bf S}_{i}\cdot {\bf S}_{i+\hat{x}-\hat{y}}\rangle] \label{sigma2s}, \\
\sigma^Q_{B2g} &=& \frac{1}{N_m}\sum_i[\langle {\bf Q}_{i}\cdot {\bf Q}_{i+\hat{x}+\hat{y}}\rangle - \langle {\bf Q}_{i}\cdot {\bf Q}_{i+\hat{x}-\hat{y}}\rangle] \label{sigma2q},
\end{eqnarray}
where $\hat{x}$ and $\hat{y}$ denote the unit vectors along the two directions, and $N_m$ is the number of sites of the two columns in the middle of cylinder. 
Both the nonzero $B_{1g}$ and $B_{2g}$ nematic order parameters characterize a $C_4$ symmetry breaking on the square lattice, but the two order parameters are related to different mirror symmetries. 
The $B_{1g}$ order preserves the mirror symmetries along $e_{x}$ and $e_{y}$ but breaks the mirror symmetries along $e_{x+y}$ and $e_{x-y}$.
On the other hand, the $B_{2g}$ order preserves the mirror symmetries along $e_{x+y}$ and $e_{x-y}$ but breaks the symmetries along $e_{x}$ and $e_{y}$. 
In DMRG simulations on cylindrical geometries, lattice $C_4$ rotational symmetry is naturally broken, and thus the corresponding nematic order parameter would be nonzero when using finite circumferences.
However, for the quantum states without $C_4$ symmetry breaking, the cylinder-induced explicit 
nonzero nematic order was found to be very small and it decreases rapidly with growing circumference (see below).
For the states with intrinsic $C_4$ symmetry breaking, on the other hand, 
a robust nematic order parameter was found. As a consequence, this aspect is under control in our study.

For the benefit of the readers, we would like to mention two additional comments: (1) we have also performed linear flavor-wave calculations~\cite{toth2010,hunsl} to study the same parameter regions as used for the DMRG simulations. However, due to the spurious divergence of both spin dipolar and quadrupolar order parameters when using this technique, we could not obtain reasonable flavor-wave phase diagrams to report. For this reason, we decided not to discuss such calculations in this publication; (2) Our investigation focuses on finding a robust region of stability of the previously reported nematic spin liquid~\cite{hunsl}. But at present we do not know which specific material may display the needed robust values of the biquadratic terms $K_1$, $K_2$, and $K_3$. This matter will be investigated in the near future using more realistic multiorbital Hubbard systems with large on-site coupling $U$.

\section{Quantum Phase Diagram with $K_2$}
\label{sec:k2}

First, we study the quantum phase diagram of the model varying $K_2$ in the absence of $K_3$, with a resulting
phase diagram shown in Fig.~\ref{pdk2}.
We found three phases in this system. 
In the nematic spin liquid around the $SU(3)$ point~\cite{hunsl}, both spin and quadrupolar structure factors in Fig.~\ref{pdk2}(e) have weak peaks at momentum $(\pi,2\pi/3)$ on the RC6 cylinder. 
As $K_2$ decreases down to $K_2<-0.2$, an antiferroquadrupolar ordered phase with dominant peak at momentum $(\pi,\pi)$ (AFQ2) arises, as shown in Figs.~\ref{pdk2}(a,d). The spin structure factor has four weak peaks at momenta $(0,\pm \pi)$ and $(\pm \pi, 0)$.
The real-space quadrupolar correlations in Fig.~\ref{pdk2}(g) clearly show two-sublattice structures in both the $x$- and $y$-directions. 
With growing $K_2>0$, the system transits to an antiferromagnetic order with dominant peak at $(\pi,2\pi/3)$ (AFM23) in the spin structure factor, as shown in Figs.~\ref{pdk2}(b,f). 
The real-space spin correlations in Fig.~\ref{pdk2}(h) also suggest two- and three-sublattice structures along the $x$- and $y$-directions, respectively. 

Considering the values of the peaks in both spin and quadrupolar structure factors, we plot the spin $(m^2_S)$ and quadrupolar $(m^2_Q)$ order parameters at different momenta as a function of $K_2$ in Figs.~\ref{j1k1k2}(a-b). 
In Fig.~\ref{j1k1k2}(a), we find a sharp drop of $m^2_S(0,\pi)/(\pi,0)$ at $K_2 \simeq -0.2$. 
Meanwhile, $m^2_Q(\pi,\pi)$ also has a sharp drop, which indicates a sharp phase transition between the AFQ2 and the nematic spin liquid, likely a first-order transition signaled by a level crossing.
For large $K_2 \simeq 1.0$, the dominant order parameter is $m^2_S(\pi,2\pi/3)$, suggesting 
the emergence of spin AFM23 order. 
Both the order parameters $m^2_S$ and $m^2_Q$ continuously change with growing $K_2 > 0.0$. 
Due to our limitations in system sizes, it is not easy to identify the precise
transition point between the nematic spin liquid and the AFM23 phase. Thus, in our sketched phase
diagram only a diffuse region is shown for the transition. 

To characterize the different ordered phases in real space, we plot log-linear figures for 
both the spin $\langle {\bf S}_i \cdot {\bf S}_j \rangle$ and quadrupolar $\langle {\bf Q}_i \cdot {\bf Q}_j \rangle$ 
correlations on the long RC6 cylinder in Figs.~\ref{j1k1k2}(c-d). 
Around the $SU(3)$ point, both spin and quadrupolar correlations decay very fast, supporting the previously reported presence of a  quantum spin liquid phase in this region~\cite{hunsl}.
For the AFQ2 phase, the quadrupolar correlation of the blue square with $K_2=-0.4$ in Fig.~\ref{j1k1k2}(d) clearly shows long-range quadrupolar order. 
In the spin correlations of Fig.~\ref{j1k1k2}(c), we find that 
the results for the purple up triangle with $K_2=1.0$ strongly suggest long-range magnetic order for the AFM23 phase. 
Also, we observe that the spin correlations at $K_2 = 0.4$ enhance dramatically compared to those for the SU(3) point and $K_2 = 0.2$, which suggests that the AFM23 order develops gradually approximately at $K_2 \sim 0.3$.

\begin{figure}[t]
\includegraphics[width = \linewidth]{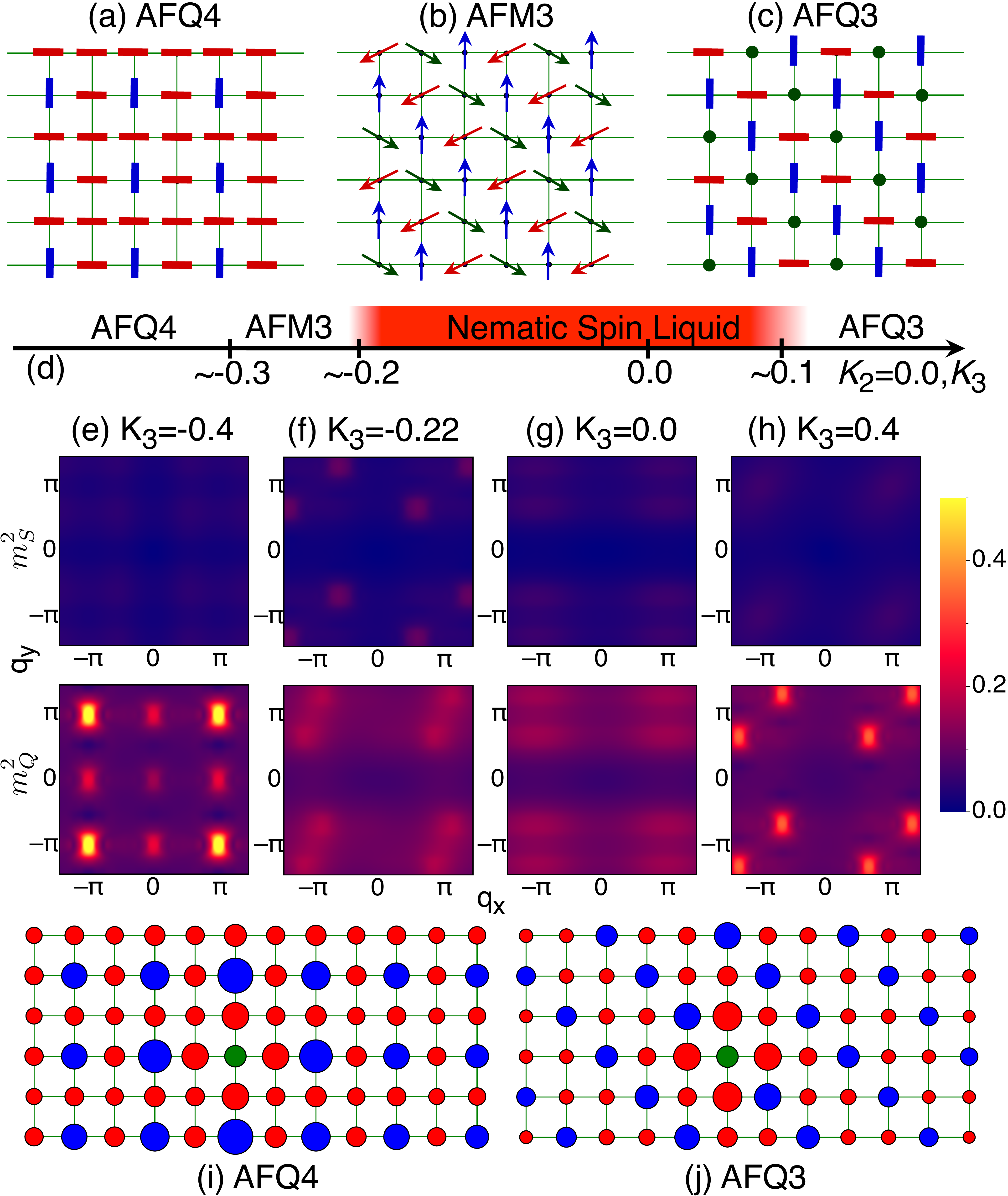}
\caption{Quantum phase diagram of the $SU(3)$ model with additional $K_3$ interaction. (a)-(c) are the sketches illustrating the AFQ4, AFM3, and AFQ3 phases, respectively. (d) Quantum phase diagram of the spin-1 $SU(3)$ Heisenberg model on the square lattice with $J_1=K_1=1.0, K_2=0.0$, and varying $K_3$. 
The regime of the nematic spin liquid around the $SU(3)$ point is indicated by red shading.
(e)-(h) are spin ($m^2_S$) and quadrupolar ($m^2_Q$) structure factors for different values of $K_3$, which are obtained from the middle $6\times 12$ sites of the RC6 cylinder. The upper and lower figures are for $m^2_S$ and $m^2_Q$, respectively. 
(i) The real-space quadrupolar correlation functions for the AFQ4 state with $K_3=-0.4$ in the middle of the RC6 cylinder. (j) The real-space quadrupolar correlation functions for the AFQ3 state with $K_3=0.4$ in the middle of the RC6 cylinder. The green site is the reference site; the blue and red colors denote positive and negative correlations of the sites with respect to the reference site, respectively. 
The areas of circles are proportional to the magnitude of correlations.}
\label{pdk3}
\end{figure}

\begin{figure}[t]
\includegraphics[width = \linewidth]{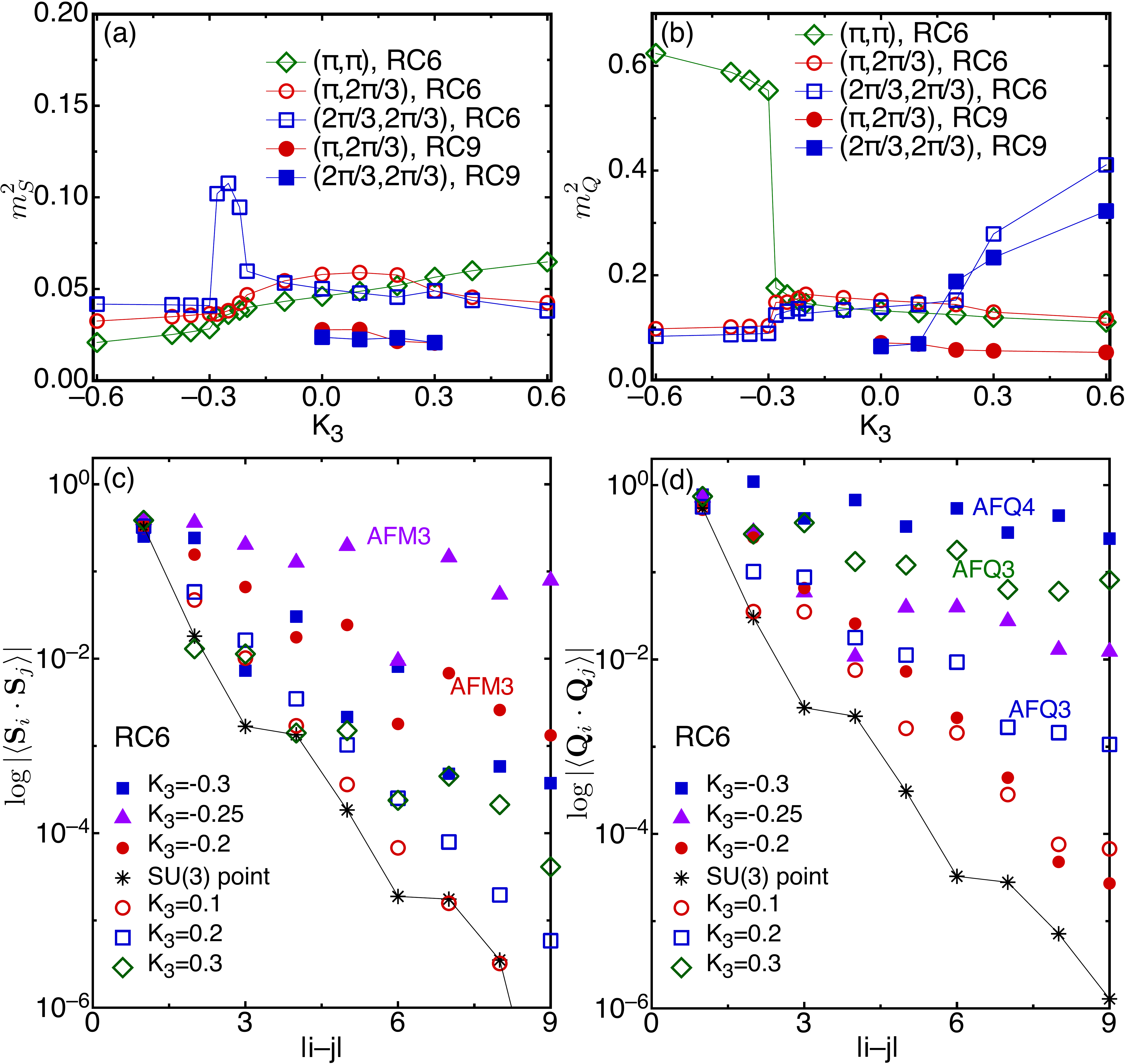}
\caption{Interaction dependence of the order parameters and correlation functions in the $SU(3)$ model with additional $K_3$ coupling. (a) and (b) are spin ($m^2_S$) and quadrupolar ($m^2_Q$) order parameters for $J_1=K_1=1.0, K_2=0.0$, at different momenta vs $K_3$, on the RC6 and RC9 cylinders. Both order parameters are obtained from the middle $6 \times 12$ sites on the RC6 cylinder or the middle $9 \times 18$ sites on the RC9 cylinder. (c) and (d) are the log-linear plots of the spin $\langle {\bf S}_i\cdot{\bf S}_j \rangle$ 
and quadrupolar $\langle {\bf Q}_i\cdot{\bf Q}_j \rangle$ 
correlations as a function of the distance on the RC6 cylinder for different values of $K_3$.}
\label{j1k1k3}
\end{figure}

\begin{figure*}[t]
\includegraphics[width = 0.9\linewidth]{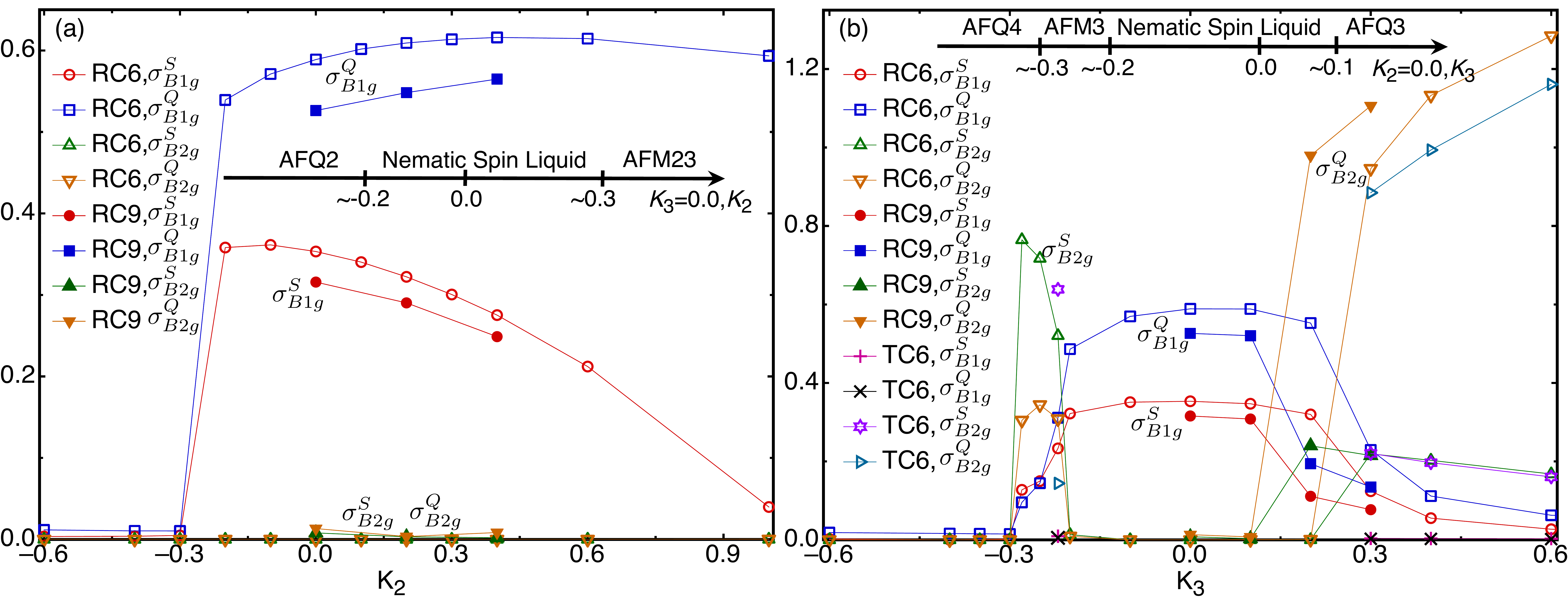}
\caption{Nematic order parameters in the $B_{1g}$ ($\sigma^S_{B1g}$ and $\sigma^Q_{B1g}$) and $B_{2g}$ ($\sigma^S_{B2g}$ and $\sigma^Q_{B2g}$) channels (a) vs $K_2$, at $K_3=0$, and (b) vs $K_3$, at $K_2 = 0$, 
on the RC6 and RC9 cylinders. The insets show the corresponding quantum phase diagrams.}
\label{sigma}
\end{figure*}

\section{Quantum Phase Diagram with $K_3$}
\label{sec:k3}

We have also investigated the model with $J_1=K_1=1.0$, varying $K_3$, at $K_2 = 0.0$. 
As in the previous case, we also obtained a rich quantum phase diagram as shown 
in Fig.~\ref{pdk3}, which has now four quantum phases. 
Besides the nematic spin liquid around the $SU(3)$ point, we also find 
three spin dipolar or quadrupolar ordered phases. 
At the $K_3 > 0.1$ region, the system displays an antiferroquadrupolar 
order with the ordering peak at momentum $(2\pi/3,2\pi/3)$ (AFQ3).
The real-space quadrupolar correlations in Fig.~\ref{pdk3}(j) show three-sublattice 
orders in both the $x$- and $y$-directions. 
For $K_3 < 0.0$, there are two quantum phase transitions. As $K_3$ decreases from $0.0$, the nematic spin liquid first transits to an antiferromagnetic order with the momentum peak at  $(2\pi/3,2\pi/3)$ (AFM3), which is stable 
only in the narrow region $-0.3<K_3<-0.2$ (the real-space spin configuration of the AFM3 is similar to the real-space quadrupolar configuration of the AFQ3). 
Note that this AFM3 state was suggested to be the ground state of the model at the $SU(3)$ point.~\cite{toth2010, toth2012,bauer2012,corboz2017,niesen2017tensor} Thus, we have the capability to observe this phase if stable.
However, in our DMRG calculation the $SU(3)$ point realizes a spin liquid state and the AFM3 state is indeed stable
but only in a neighboring region, which implies that the AFM3 state is indeed a very competitive state in the search for the ground state of the $SU(3)$ model.
The second phase transition happens at $K_3 \simeq -0.3$, where the system moves into an antiferroquadrupolar phase with dominant peak at momentum $(\pi,\pi)$ and weak peaks at momenta $(\pi,0)$, $(0,\pi), (0,0)$ (AFQ4). 
According to the real-space quadrupolar correlations shown in Fig.~\ref{pdk3}(i), the unit cell of the AFQ4 state includes four sites.

In Figs.~\ref{j1k1k3}(a-b), we show the spin ($m^2_S$) and quadrupolar ($m^2_Q$) order 
parameters at different momenta as a function of $K_3$. 
With growing $K_3$ starting from $K_3 = -0.6$, the magnetic order parameter $m^2_S(2\pi/3, 2\pi/3)$ has a sharp jump at $K_3 \simeq -0.3$ and there is a sharp drop of the quadrupolar order parameter $m^2_Q(\pi, \pi)$ as well, which indicates a sharp phase transition (first-order level crossing) between the AFQ4 and AFM3 phases. 
The AFM3 phase exists in a small region, and $m^2_S(2\pi/3, 2\pi/3)$ drops sharply at $K_3 \simeq -0.2$, suggesting again a sharp transition to the nematic spin liquid. 
The third phase transition happens at $K_3 \simeq 0.1$ between the nematic spin liquid and the AFQ3 phase, which is characterized by an increasing $m^2_Q(2\pi/3, 2\pi/3)$. 

The spin and quadrupolar order phases are further characterized by spin and quadrupolar correlations in the long RC6 cylinder, as shown in Figs.~\ref{j1k1k3}(c-d). 
Although the AFM3 phase region is narrow, the AFM3 order is clearly shown at $K_3=-0.25$ by the 
spin correlations that decay quite slowly.
A larger region of stability for the AFM3 order might be realized by considering other interactions.
For $K_3=-0.3$ and $0.3$, the AFQ4 and the AFQ3 orders are supported by strong quadrupolar correlations.

\section{Nematicity}
\label{sec:nematic}

Since some of the spin dipolar and quadrupolar orders also break lattice rotational symmetry, we now turn to analyzing the nematicity of the different quantum phases by calculating four nematic order parameters in both the $B_{1g}$ and $B_{2g}$ channels for spin ($\sigma^S_{B1g/B2g}$) and quadrupolar ($\sigma^Q_{B1g/B2g}$) operators, as defined in Eqs.\eqref{sigma1s}-\eqref{sigma2q}. 
The nematic order parameters for $J_1=K_1=1.0, K_3=0.0$, as a function of $K_2$ are shown in Fig.~\ref{sigma}(a).
In the AFQ2 phase which preserves $C_4$ symmetry, both the $B_{1g}$ and $B_{2g}$ nematic orders in spin and quadrupolar channels are quite small. The fact that the $B_{1g}$ nematicity is
not exactly zero (although it is very small) in the AFQ2 regime 
is because of the explicit symmetry breaking caused by the cylindrical geometry
in this channel (while in the $B_{2g}$ channel the cylinder does not explicitly break the symmetry). The $B_{1g}$ quantities 
should decay to zero as the width of the cylinder increases in the AFQ2 phase. The results 
also indicate that the symmetry-breaking effects due to the cylindrical geometry are small.
As $K_2$ increases from negative values, both the $B_{1g}$ nematic order parameters $\sigma^S_{B1g}$ and $\sigma^Q_{B1g}$ present a big jump at $K_2 \simeq -0.2$, which is consistent with the sharp phase transition from the AFQ2 to the nematic spin liquid with $C_4$ symmetry breaking discussed before.
With further growing $K_2$, the $B_{2g}$ nematic orders remain quite small in the spin liquid but 
the $B_{1g}$ orders are finite, as recently reported, which indicate that both the nematic spin liquid 
and the AFM23 states have the mirror symmetries along $e_{x}$ and $e_{y}$ but break the mirror symmetries 
along $e_{x+y}$ and $e_{x-y}$. 
We wish to remark that in both the nematic spin liquid as well as the AFM23 phase stabilized at larger $K_2$, the quadrupolar nematicity in the $B_{1g}$ $\sigma^Q_{B1g}$ is dominant, especially deep in the AFM23 phase such as $K_2=1.0$. 

For the case with $J_1=K_1=1.0, K_2=0.0$, and changing $K_3$, we show the nematic orders in Fig.~\ref{sigma}(b). 
Similarly to the AFQ2 phase, both the $B_{1g}$ and $B_{2g}$ nematic orders are small for the AFQ4 phase with $C_4$ symmetry. 
With further growing $K_3$, we find three sharp changes of the $B_{2g}$ nematic orders at $K_3 \simeq -0.3, -0.2$, 
and $0.1$, which agrees with the three phase transitions shown in Fig.~\ref{pdk3}. 
In the AFM3 phase, the $B_{2g}$ nematicity in the spin channel $\sigma^S_{B2g}$ is dominant, while $\sigma^Q_{B2g}$ is larger in the AFQ3 phase. 
The strong $B_{2g}$ nematicity indicates that the AFM3 and AFQ3 orders break the mirror symmetries along $e_{x}$ and $e_{y}$. We noticed that there are small values of $\sigma^S_{B1g}$ and $\sigma^Q_{B1g}$ in both the AFM3 and AFQ3 phases, which decay in magnitude from RC6 to RC9 and will likely vanish in the thermodynamic limit. Furthermore, we have performed DMRG simulations for the AFM3 and AFQ3 phases on twisted TC6 cylinders, which are rotated by the RC6 geometry by $45^{\circ}$. In this case we obtain much smaller $B_{1g}$ nematic order parameters as shown in Fig.~\ref{sigma}, which is consistent with our overall description of results.

\section{Summary and Discussion}
\label{sec:sum}

In this paper we have studied the quantum phases and phase transitions in the vicinity of the highly symmetric $SU(3)$ point in the spin-$1$ bilinear-biquadratic Heisenberg model on the square lattice by using large-scale DMRG simulation. 
By computing spin and quadrupolar order parameters and correlation functions with tuning either the second-neighbor ($K_2$) or the third-neighbor ($K_3$) biquadratic interaction, we establish rich quantum phase diagrams with different spin dipolar and quadrupolar ordered phases.
In addition, the nematic spin liquid phase which has been reported recently near the $SU(3)$ model~\cite{hunsl} is found to be robust in a wide parameter region varying the $K_2, K_3$ couplings.
Furthermore, we calculate the nematic order parameters for different quantum phases. Nematic magnetic states break the lattice
invariance under 90$^o$ rotations.
We identify different lattice nematicity channels in the many quantum phases unveiled 
by using the $B_{1g}$ and $B_{2g}$ nematic orders.
We also observe that the quantum phase transitions characterized by the nematic order parameters are consistent with those we find by spin and quadrupolar order parameters.

\section*{Acknowledgment}
We thank A. H. Nevidomskyy, H. Hu, R. Yu, and C. Batista for fruitful discussions.
E.D. and W.-J.H. were supported by the U.S. Department of Energy (DOE), Office of Science, Basic Energy Sciences (BES),  Materials  Science  and  Engineering  Division.
S.S.G. was supported by NSFC grants No. 11874078, 11834014 and the Fundamental Research Funds for the Central Universities.
H.-H.L. and Q.S. were supported by the U.S. Department of Energy, Office of Science, Basic Energy Sciences, under Award No.\ DE-SC0018197, and the Robert A.\ Welch Foundation Grant No.\ C-1411.
The majority of the computational calculations have been performed on the Extreme Science and Engineering Discovery Environment (XSEDE) supported by NSF under Grant No.\ DMR160057. Most of the DMRG calculations have been done by W.-J.H. and H.-H.L. while at Rice University.

\bibliographystyle{apsrev}
\bibliography{spin1pd}{}

\begin{thebibliography}{60}
\expandafter\ifx\csname natexlab\endcsname\relax\def\natexlab#1{#1}\fi
\expandafter\ifx\csname bibnamefont\endcsname\relax
  \def\bibnamefont#1{#1}\fi
\expandafter\ifx\csname bibfnamefont\endcsname\relax
  \def\bibfnamefont#1{#1}\fi
\expandafter\ifx\csname citenamefont\endcsname\relax
  \def\citenamefont#1{#1}\fi
\expandafter\ifx\csname url\endcsname\relax
  \def\url#1{\texttt{#1}}\fi
\expandafter\ifx\csname urlprefix\endcsname\relax\def\urlprefix{URL }\fi
\providecommand{\bibinfo}[2]{#2}
\providecommand{\eprint}[2][]{\url{#2}}

\bibitem[{\citenamefont{Musaelian and Joynt}(1996)}]{musaelian1996}
\bibinfo{author}{\bibfnamefont{K.}~\bibnamefont{Musaelian}} \bibnamefont{and}
  \bibinfo{author}{\bibfnamefont{R.}~\bibnamefont{Joynt}},
  \bibinfo{journal}{Journal of Physics: Condensed Matter}
  \textbf{\bibinfo{volume}{8}}, \bibinfo{pages}{L105} (\bibinfo{year}{1996}),
  \urlprefix\url{https://doi.org/10.1088%2F0953-8984%2F8%2F8%2F002}.

\bibitem[{\citenamefont{Balents}(1996)}]{balents1996}
\bibinfo{author}{\bibfnamefont{L.}~\bibnamefont{Balents}},
  \bibinfo{journal}{EPL (Europhysics Letters)} \textbf{\bibinfo{volume}{33}},
  \bibinfo{pages}{291} (\bibinfo{year}{1996}),
  \urlprefix\url{https://doi.org/10.1209%2Fepl%2Fi1996-00335-x}.

\bibitem[{\citenamefont{Mulligan et~al.}(2010)\citenamefont{Mulligan, Nayak,
  and Kachru}}]{Mulligan2010}
\bibinfo{author}{\bibfnamefont{M.}~\bibnamefont{Mulligan}},
  \bibinfo{author}{\bibfnamefont{C.}~\bibnamefont{Nayak}}, \bibnamefont{and}
  \bibinfo{author}{\bibfnamefont{S.}~\bibnamefont{Kachru}},
  \bibinfo{journal}{Phys. Rev. B} \textbf{\bibinfo{volume}{82}},
  \bibinfo{pages}{085102} (\bibinfo{year}{2010}),
  \urlprefix\url{https://link.aps.org/doi/10.1103/PhysRevB.82.085102}.

\bibitem[{\citenamefont{Mulligan et~al.}(2011)\citenamefont{Mulligan, Nayak,
  and Kachru}}]{Mulligan2011}
\bibinfo{author}{\bibfnamefont{M.}~\bibnamefont{Mulligan}},
  \bibinfo{author}{\bibfnamefont{C.}~\bibnamefont{Nayak}}, \bibnamefont{and}
  \bibinfo{author}{\bibfnamefont{S.}~\bibnamefont{Kachru}},
  \bibinfo{journal}{Phys. Rev. B} \textbf{\bibinfo{volume}{84}},
  \bibinfo{pages}{195124} (\bibinfo{year}{2011}),
  \urlprefix\url{https://link.aps.org/doi/10.1103/PhysRevB.84.195124}.

\bibitem[{\citenamefont{Maciejko et~al.}(2013)\citenamefont{Maciejko, Hsu,
  Kivelson, Park, and Sondhi}}]{Maciejko2013}
\bibinfo{author}{\bibfnamefont{J.}~\bibnamefont{Maciejko}},
  \bibinfo{author}{\bibfnamefont{B.}~\bibnamefont{Hsu}},
  \bibinfo{author}{\bibfnamefont{S.~A.} \bibnamefont{Kivelson}},
  \bibinfo{author}{\bibfnamefont{Y.}~\bibnamefont{Park}}, \bibnamefont{and}
  \bibinfo{author}{\bibfnamefont{S.~L.} \bibnamefont{Sondhi}},
  \bibinfo{journal}{Phys. Rev. B} \textbf{\bibinfo{volume}{88}},
  \bibinfo{pages}{125137} (\bibinfo{year}{2013}),
  \urlprefix\url{https://link.aps.org/doi/10.1103/PhysRevB.88.125137}.

\bibitem[{\citenamefont{Liu et~al.}(2013)\citenamefont{Liu, Hasdemir, Shayegan,
  Pfeiffer, West, and Baldwin}}]{Liu2013}
\bibinfo{author}{\bibfnamefont{Y.}~\bibnamefont{Liu}},
  \bibinfo{author}{\bibfnamefont{S.}~\bibnamefont{Hasdemir}},
  \bibinfo{author}{\bibfnamefont{M.}~\bibnamefont{Shayegan}},
  \bibinfo{author}{\bibfnamefont{L.~N.} \bibnamefont{Pfeiffer}},
  \bibinfo{author}{\bibfnamefont{K.~W.} \bibnamefont{West}}, \bibnamefont{and}
  \bibinfo{author}{\bibfnamefont{K.~W.} \bibnamefont{Baldwin}},
  \bibinfo{journal}{Phys. Rev. B} \textbf{\bibinfo{volume}{88}},
  \bibinfo{pages}{035307} (\bibinfo{year}{2013}),
  \urlprefix\url{https://link.aps.org/doi/10.1103/PhysRevB.88.035307}.

\bibitem[{\citenamefont{You et~al.}(2014)\citenamefont{You, Cho, and
  Fradkin}}]{You2014}
\bibinfo{author}{\bibfnamefont{Y.}~\bibnamefont{You}},
  \bibinfo{author}{\bibfnamefont{G.~Y.} \bibnamefont{Cho}}, \bibnamefont{and}
  \bibinfo{author}{\bibfnamefont{E.}~\bibnamefont{Fradkin}},
  \bibinfo{journal}{Phys. Rev. X} \textbf{\bibinfo{volume}{4}},
  \bibinfo{pages}{041050} (\bibinfo{year}{2014}),
  \urlprefix\url{https://link.aps.org/doi/10.1103/PhysRevX.4.041050}.

\bibitem[{\citenamefont{Gromov and Son}(2017)}]{Gromov2017}
\bibinfo{author}{\bibfnamefont{A.}~\bibnamefont{Gromov}} \bibnamefont{and}
  \bibinfo{author}{\bibfnamefont{D.~T.} \bibnamefont{Son}},
  \bibinfo{journal}{Phys. Rev. X} \textbf{\bibinfo{volume}{7}},
  \bibinfo{pages}{041032} (\bibinfo{year}{2017}),
  \urlprefix\url{https://link.aps.org/doi/10.1103/PhysRevX.7.041032}.

\bibitem[{\citenamefont{Du et~al.}(2019)\citenamefont{Du, Wurstbauer, West,
  Pfeiffer, Fallahi, Gardner, Manfra, and Pinczuk}}]{du2019}
\bibinfo{author}{\bibfnamefont{L.}~\bibnamefont{Du}},
  \bibinfo{author}{\bibfnamefont{U.}~\bibnamefont{Wurstbauer}},
  \bibinfo{author}{\bibfnamefont{K.~W.} \bibnamefont{West}},
  \bibinfo{author}{\bibfnamefont{L.~N.} \bibnamefont{Pfeiffer}},
  \bibinfo{author}{\bibfnamefont{S.}~\bibnamefont{Fallahi}},
  \bibinfo{author}{\bibfnamefont{G.~C.} \bibnamefont{Gardner}},
  \bibinfo{author}{\bibfnamefont{M.~J.} \bibnamefont{Manfra}},
  \bibnamefont{and} \bibinfo{author}{\bibfnamefont{A.}~\bibnamefont{Pinczuk}},
  \bibinfo{journal}{Science Advances} \textbf{\bibinfo{volume}{5}},
  \bibinfo{pages}{eaav3407} (\bibinfo{year}{2019}),
  \urlprefix\url{https://advances.sciencemag.org/content/5/3/eaav3407}.

\bibitem[{\citenamefont{Fang et~al.}(2008)\citenamefont{Fang, Yao, Tsai, Hu,
  and Kivelson}}]{Kivelson2008}
\bibinfo{author}{\bibfnamefont{C.}~\bibnamefont{Fang}},
  \bibinfo{author}{\bibfnamefont{H.}~\bibnamefont{Yao}},
  \bibinfo{author}{\bibfnamefont{W.-F.} \bibnamefont{Tsai}},
  \bibinfo{author}{\bibfnamefont{J.}~\bibnamefont{Hu}}, \bibnamefont{and}
  \bibinfo{author}{\bibfnamefont{S.~A.} \bibnamefont{Kivelson}},
  \bibinfo{journal}{Phys. Rev. B} \textbf{\bibinfo{volume}{77}},
  \bibinfo{pages}{224509} (\bibinfo{year}{2008}),
  \urlprefix\url{https://link.aps.org/doi/10.1103/PhysRevB.77.224509}.

\bibitem[{\citenamefont{Xu et~al.}(2008)\citenamefont{Xu, M\"uller, and
  Sachdev}}]{Sachdev2008}
\bibinfo{author}{\bibfnamefont{C.}~\bibnamefont{Xu}},
  \bibinfo{author}{\bibfnamefont{M.}~\bibnamefont{M\"uller}}, \bibnamefont{and}
  \bibinfo{author}{\bibfnamefont{S.}~\bibnamefont{Sachdev}},
  \bibinfo{journal}{Phys. Rev. B} \textbf{\bibinfo{volume}{78}},
  \bibinfo{pages}{020501} (\bibinfo{year}{2008}),
  \urlprefix\url{https://link.aps.org/doi/10.1103/PhysRevB.78.020501}.

\bibitem[{\citenamefont{Bao et~al.}(2009)\citenamefont{Bao, Qiu, Huang, Green,
  Zajdel, Fitzsimmons, Zhernenkov, Chang, Fang, Qian et~al.}}]{bao2009}
\bibinfo{author}{\bibfnamefont{W.}~\bibnamefont{Bao}},
  \bibinfo{author}{\bibfnamefont{Y.}~\bibnamefont{Qiu}},
  \bibinfo{author}{\bibfnamefont{Q.}~\bibnamefont{Huang}},
  \bibinfo{author}{\bibfnamefont{M.~A.} \bibnamefont{Green}},
  \bibinfo{author}{\bibfnamefont{P.}~\bibnamefont{Zajdel}},
  \bibinfo{author}{\bibfnamefont{M.~R.} \bibnamefont{Fitzsimmons}},
  \bibinfo{author}{\bibfnamefont{M.}~\bibnamefont{Zhernenkov}},
  \bibinfo{author}{\bibfnamefont{S.}~\bibnamefont{Chang}},
  \bibinfo{author}{\bibfnamefont{M.}~\bibnamefont{Fang}},
  \bibinfo{author}{\bibfnamefont{B.}~\bibnamefont{Qian}}, \bibnamefont{et~al.},
  \bibinfo{journal}{Phys. Rev. Lett.} \textbf{\bibinfo{volume}{102}},
  \bibinfo{pages}{247001} (\bibinfo{year}{2009}),
  \urlprefix\url{https://link.aps.org/doi/10.1103/PhysRevLett.102.247001}.

\bibitem[{\citenamefont{Li et~al.}(2009)\citenamefont{Li, de~la Cruz, Huang,
  Chen, Lynn, Hu, Huang, Hsu, Yeh, Wu et~al.}}]{Shiliang2009}
\bibinfo{author}{\bibfnamefont{S.}~\bibnamefont{Li}},
  \bibinfo{author}{\bibfnamefont{C.}~\bibnamefont{de~la Cruz}},
  \bibinfo{author}{\bibfnamefont{Q.}~\bibnamefont{Huang}},
  \bibinfo{author}{\bibfnamefont{Y.}~\bibnamefont{Chen}},
  \bibinfo{author}{\bibfnamefont{J.~W.} \bibnamefont{Lynn}},
  \bibinfo{author}{\bibfnamefont{J.}~\bibnamefont{Hu}},
  \bibinfo{author}{\bibfnamefont{Y.-L.} \bibnamefont{Huang}},
  \bibinfo{author}{\bibfnamefont{F.-C.} \bibnamefont{Hsu}},
  \bibinfo{author}{\bibfnamefont{K.-W.} \bibnamefont{Yeh}},
  \bibinfo{author}{\bibfnamefont{M.-K.} \bibnamefont{Wu}},
  \bibnamefont{et~al.}, \bibinfo{journal}{Phys. Rev. B}
  \textbf{\bibinfo{volume}{79}}, \bibinfo{pages}{054503}
  (\bibinfo{year}{2009}),
  \urlprefix\url{https://link.aps.org/doi/10.1103/PhysRevB.79.054503}.

\bibitem[{\citenamefont{Dai et~al.}(2009)\citenamefont{Dai, Si, Zhu, and
  Abrahams}}]{Dai4118}
\bibinfo{author}{\bibfnamefont{J.}~\bibnamefont{Dai}},
  \bibinfo{author}{\bibfnamefont{Q.}~\bibnamefont{Si}},
  \bibinfo{author}{\bibfnamefont{J.-X.} \bibnamefont{Zhu}}, \bibnamefont{and}
  \bibinfo{author}{\bibfnamefont{E.}~\bibnamefont{Abrahams}},
  \bibinfo{journal}{Proceedings of the National Academy of Sciences}
  \textbf{\bibinfo{volume}{106}}, \bibinfo{pages}{4118} (\bibinfo{year}{2009}),
  \urlprefix\url{https://www.pnas.org/content/106/11/4118}.

\bibitem[{\citenamefont{Chu et~al.}(2010)\citenamefont{Chu, Analytis, De~Greve,
  McMahon, Islam, Yamamoto, and Fisher}}]{Chu824}
\bibinfo{author}{\bibfnamefont{J.-H.} \bibnamefont{Chu}},
  \bibinfo{author}{\bibfnamefont{J.~G.} \bibnamefont{Analytis}},
  \bibinfo{author}{\bibfnamefont{K.}~\bibnamefont{De~Greve}},
  \bibinfo{author}{\bibfnamefont{P.~L.} \bibnamefont{McMahon}},
  \bibinfo{author}{\bibfnamefont{Z.}~\bibnamefont{Islam}},
  \bibinfo{author}{\bibfnamefont{Y.}~\bibnamefont{Yamamoto}}, \bibnamefont{and}
  \bibinfo{author}{\bibfnamefont{I.~R.} \bibnamefont{Fisher}},
  \bibinfo{journal}{Science} \textbf{\bibinfo{volume}{329}},
  \bibinfo{pages}{824} (\bibinfo{year}{2010}),
  \urlprefix\url{https://science.sciencemag.org/content/329/5993/824}.

\bibitem[{\citenamefont{Dai et~al.}(2012)\citenamefont{Dai, Hu, and
  Dagotto}}]{dai2012magnetism}
\bibinfo{author}{\bibfnamefont{P.}~\bibnamefont{Dai}},
  \bibinfo{author}{\bibfnamefont{J.}~\bibnamefont{Hu}}, \bibnamefont{and}
  \bibinfo{author}{\bibfnamefont{E.}~\bibnamefont{Dagotto}},
  \bibinfo{journal}{Nature Physics} \textbf{\bibinfo{volume}{8}},
  \bibinfo{pages}{709} (\bibinfo{year}{2012}),
  \urlprefix\url{https://doi.org/10.1038/nphys2438}.

\bibitem[{\citenamefont{Liang et~al.}(2013)\citenamefont{Liang, Moreo, and
  Dagotto}}]{liang2013}
\bibinfo{author}{\bibfnamefont{S.}~\bibnamefont{Liang}},
  \bibinfo{author}{\bibfnamefont{A.}~\bibnamefont{Moreo}}, \bibnamefont{and}
  \bibinfo{author}{\bibfnamefont{E.}~\bibnamefont{Dagotto}},
  \bibinfo{journal}{Phys. Rev. Lett.} \textbf{\bibinfo{volume}{111}},
  \bibinfo{pages}{047004} (\bibinfo{year}{2013}),
  \urlprefix\url{https://link.aps.org/doi/10.1103/PhysRevLett.111.047004}.

\bibitem[{\citenamefont{Fernandes et~al.}(2014)\citenamefont{Fernandes,
  Chubukov, and Schmalian}}]{fernandes2014drives}
\bibinfo{author}{\bibfnamefont{R.~M.} \bibnamefont{Fernandes}},
  \bibinfo{author}{\bibfnamefont{A.~V.} \bibnamefont{Chubukov}},
  \bibnamefont{and}
  \bibinfo{author}{\bibfnamefont{J.}~\bibnamefont{Schmalian}},
  \bibinfo{journal}{Nature physics} \textbf{\bibinfo{volume}{10}},
  \bibinfo{pages}{97} (\bibinfo{year}{2014}),
  \urlprefix\url{https://doi.org/10.1038/nphys2877}.

\bibitem[{\citenamefont{Lu et~al.}(2014)\citenamefont{Lu, Park, Zhang, Luo,
  Nevidomskyy, Si, and Dai}}]{Lu657}
\bibinfo{author}{\bibfnamefont{X.}~\bibnamefont{Lu}},
  \bibinfo{author}{\bibfnamefont{J.~T.} \bibnamefont{Park}},
  \bibinfo{author}{\bibfnamefont{R.}~\bibnamefont{Zhang}},
  \bibinfo{author}{\bibfnamefont{H.}~\bibnamefont{Luo}},
  \bibinfo{author}{\bibfnamefont{A.~H.} \bibnamefont{Nevidomskyy}},
  \bibinfo{author}{\bibfnamefont{Q.}~\bibnamefont{Si}}, \bibnamefont{and}
  \bibinfo{author}{\bibfnamefont{P.}~\bibnamefont{Dai}},
  \bibinfo{journal}{Science} \textbf{\bibinfo{volume}{345}},
  \bibinfo{pages}{657} (\bibinfo{year}{2014}),
  \urlprefix\url{https://science.sciencemag.org/content/345/6197/657}.

\bibitem[{\citenamefont{Bishop et~al.}(2016)\citenamefont{Bishop, Moreo, and
  Dagotto}}]{Bishop2016}
\bibinfo{author}{\bibfnamefont{C.~B.} \bibnamefont{Bishop}},
  \bibinfo{author}{\bibfnamefont{A.}~\bibnamefont{Moreo}}, \bibnamefont{and}
  \bibinfo{author}{\bibfnamefont{E.}~\bibnamefont{Dagotto}},
  \bibinfo{journal}{Phys. Rev. Lett.} \textbf{\bibinfo{volume}{117}},
  \bibinfo{pages}{117201} (\bibinfo{year}{2016}),
  \urlprefix\url{https://link.aps.org/doi/10.1103/PhysRevLett.117.117201}.

\bibitem[{\citenamefont{Bishop et~al.}(2017)\citenamefont{Bishop, Herbrych,
  Dagotto, and Moreo}}]{Bishop2017}
\bibinfo{author}{\bibfnamefont{C.~B.} \bibnamefont{Bishop}},
  \bibinfo{author}{\bibfnamefont{J.}~\bibnamefont{Herbrych}},
  \bibinfo{author}{\bibfnamefont{E.}~\bibnamefont{Dagotto}}, \bibnamefont{and}
  \bibinfo{author}{\bibfnamefont{A.}~\bibnamefont{Moreo}},
  \bibinfo{journal}{Phys. Rev. B} \textbf{\bibinfo{volume}{96}},
  \bibinfo{pages}{035144} (\bibinfo{year}{2017}),
  \urlprefix\url{https://link.aps.org/doi/10.1103/PhysRevB.96.035144}.

\bibitem[{\citenamefont{Hu et~al.}(2019)\citenamefont{Hu, Gong, Lai, Hu, Si,
  and Nevidomskyy}}]{hunsl}
\bibinfo{author}{\bibfnamefont{W.-J.} \bibnamefont{Hu}},
  \bibinfo{author}{\bibfnamefont{S.-S.} \bibnamefont{Gong}},
  \bibinfo{author}{\bibfnamefont{H.-H.} \bibnamefont{Lai}},
  \bibinfo{author}{\bibfnamefont{H.}~\bibnamefont{Hu}},
  \bibinfo{author}{\bibfnamefont{Q.}~\bibnamefont{Si}}, \bibnamefont{and}
  \bibinfo{author}{\bibfnamefont{A.~H.} \bibnamefont{Nevidomskyy}},
  \bibinfo{journal}{Phys. Rev. B} \textbf{\bibinfo{volume}{100}},
  \bibinfo{pages}{165142} (\bibinfo{year}{2019}),
  \urlprefix\url{https://link.aps.org/doi/10.1103/PhysRevB.100.165142}.

\bibitem[{\citenamefont{Papanicolaou}(1988)}]{papanicolaou1988}
\bibinfo{author}{\bibfnamefont{N.}~\bibnamefont{Papanicolaou}},
  \bibinfo{journal}{Nuclear Physics B} \textbf{\bibinfo{volume}{305}},
  \bibinfo{pages}{367} (\bibinfo{year}{1988}),
  \urlprefix\url{https://doi.org/10.1016/0550-3213(88)90073-9}.

\bibitem[{\citenamefont{Nakatsuji et~al.}(2005)\citenamefont{Nakatsuji, Nambu,
  Tonomura, Sakai, Jonas, Broholm, Tsunetsugu, Qiu, and Maeno}}]{nakatsuji2005}
\bibinfo{author}{\bibfnamefont{S.}~\bibnamefont{Nakatsuji}},
  \bibinfo{author}{\bibfnamefont{Y.}~\bibnamefont{Nambu}},
  \bibinfo{author}{\bibfnamefont{H.}~\bibnamefont{Tonomura}},
  \bibinfo{author}{\bibfnamefont{O.}~\bibnamefont{Sakai}},
  \bibinfo{author}{\bibfnamefont{S.}~\bibnamefont{Jonas}},
  \bibinfo{author}{\bibfnamefont{C.}~\bibnamefont{Broholm}},
  \bibinfo{author}{\bibfnamefont{H.}~\bibnamefont{Tsunetsugu}},
  \bibinfo{author}{\bibfnamefont{Y.}~\bibnamefont{Qiu}}, \bibnamefont{and}
  \bibinfo{author}{\bibfnamefont{Y.}~\bibnamefont{Maeno}},
  \bibinfo{journal}{Science} \textbf{\bibinfo{volume}{309}},
  \bibinfo{pages}{1697} (\bibinfo{year}{2005}),
  \urlprefix\url{https://science.sciencemag.org/content/309/5741/1697}.

\bibitem[{\citenamefont{Cheng et~al.}(2011)\citenamefont{Cheng, Li, Balicas,
  Zhou, Goodenough, Xu, and Zhou}}]{cheng2011}
\bibinfo{author}{\bibfnamefont{J.~G.} \bibnamefont{Cheng}},
  \bibinfo{author}{\bibfnamefont{G.}~\bibnamefont{Li}},
  \bibinfo{author}{\bibfnamefont{L.}~\bibnamefont{Balicas}},
  \bibinfo{author}{\bibfnamefont{J.~S.} \bibnamefont{Zhou}},
  \bibinfo{author}{\bibfnamefont{J.~B.} \bibnamefont{Goodenough}},
  \bibinfo{author}{\bibfnamefont{C.}~\bibnamefont{Xu}}, \bibnamefont{and}
  \bibinfo{author}{\bibfnamefont{H.~D.} \bibnamefont{Zhou}},
  \bibinfo{journal}{Phys. Rev. Lett.} \textbf{\bibinfo{volume}{107}},
  \bibinfo{pages}{197204} (\bibinfo{year}{2011}),
  \urlprefix\url{http://link.aps.org/doi/10.1103/PhysRevLett.107.197204}.

\bibitem[{\citenamefont{F\aa{}k et~al.}(2017)\citenamefont{F\aa{}k, Bieri,
  Can\'evet, Messio, Payen, Viaud, Guillot-Deudon, Darie, Ollivier, and
  Mendels}}]{fak2017}
\bibinfo{author}{\bibfnamefont{B.}~\bibnamefont{F\aa{}k}},
  \bibinfo{author}{\bibfnamefont{S.}~\bibnamefont{Bieri}},
  \bibinfo{author}{\bibfnamefont{E.}~\bibnamefont{Can\'evet}},
  \bibinfo{author}{\bibfnamefont{L.}~\bibnamefont{Messio}},
  \bibinfo{author}{\bibfnamefont{C.}~\bibnamefont{Payen}},
  \bibinfo{author}{\bibfnamefont{M.}~\bibnamefont{Viaud}},
  \bibinfo{author}{\bibfnamefont{C.}~\bibnamefont{Guillot-Deudon}},
  \bibinfo{author}{\bibfnamefont{C.}~\bibnamefont{Darie}},
  \bibinfo{author}{\bibfnamefont{J.}~\bibnamefont{Ollivier}}, \bibnamefont{and}
  \bibinfo{author}{\bibfnamefont{P.}~\bibnamefont{Mendels}},
  \bibinfo{journal}{Phys. Rev. B} \textbf{\bibinfo{volume}{95}},
  \bibinfo{pages}{060402} (\bibinfo{year}{2017}),
  \urlprefix\url{http://link.aps.org/doi/10.1103/PhysRevB.95.060402}.

\bibitem[{\citenamefont{Quilliam et~al.}(2016)\citenamefont{Quilliam, Bert,
  Manseau, Darie, Guillot-Deudon, Payen, Baines, Amato, and
  Mendels}}]{quilliam2016}
\bibinfo{author}{\bibfnamefont{J.~A.} \bibnamefont{Quilliam}},
  \bibinfo{author}{\bibfnamefont{F.}~\bibnamefont{Bert}},
  \bibinfo{author}{\bibfnamefont{A.}~\bibnamefont{Manseau}},
  \bibinfo{author}{\bibfnamefont{C.}~\bibnamefont{Darie}},
  \bibinfo{author}{\bibfnamefont{C.}~\bibnamefont{Guillot-Deudon}},
  \bibinfo{author}{\bibfnamefont{C.}~\bibnamefont{Payen}},
  \bibinfo{author}{\bibfnamefont{C.}~\bibnamefont{Baines}},
  \bibinfo{author}{\bibfnamefont{A.}~\bibnamefont{Amato}}, \bibnamefont{and}
  \bibinfo{author}{\bibfnamefont{P.}~\bibnamefont{Mendels}},
  \bibinfo{journal}{Phys. Rev. B} \textbf{\bibinfo{volume}{93}},
  \bibinfo{pages}{214432} (\bibinfo{year}{2016}),
  \urlprefix\url{http://link.aps.org/doi/10.1103/PhysRevB.93.214432}.

\bibitem[{\citenamefont{Yu et~al.}(2012)\citenamefont{Yu, Wang, Goswami,
  Nevidomskyy, Si, and Abrahams}}]{rong2012}
\bibinfo{author}{\bibfnamefont{R.}~\bibnamefont{Yu}},
  \bibinfo{author}{\bibfnamefont{Z.}~\bibnamefont{Wang}},
  \bibinfo{author}{\bibfnamefont{P.}~\bibnamefont{Goswami}},
  \bibinfo{author}{\bibfnamefont{A.~H.} \bibnamefont{Nevidomskyy}},
  \bibinfo{author}{\bibfnamefont{Q.}~\bibnamefont{Si}}, \bibnamefont{and}
  \bibinfo{author}{\bibfnamefont{E.}~\bibnamefont{Abrahams}},
  \bibinfo{journal}{Phys. Rev. B} \textbf{\bibinfo{volume}{86}},
  \bibinfo{pages}{085148} (\bibinfo{year}{2012}),
  \urlprefix\url{https://link.aps.org/doi/10.1103/PhysRevB.86.085148}.

\bibitem[{\citenamefont{Yu and Si}(2015)}]{rong2015}
\bibinfo{author}{\bibfnamefont{R.}~\bibnamefont{Yu}} \bibnamefont{and}
  \bibinfo{author}{\bibfnamefont{Q.}~\bibnamefont{Si}}, \bibinfo{journal}{Phys.
  Rev. Lett.} \textbf{\bibinfo{volume}{115}}, \bibinfo{pages}{116401}
  (\bibinfo{year}{2015}),
  \urlprefix\url{https://link.aps.org/doi/10.1103/PhysRevLett.115.116401}.

\bibitem[{\citenamefont{Wang et~al.}(2016)\citenamefont{Wang, Hu, and
  Nevidomskyy}}]{wang2016}
\bibinfo{author}{\bibfnamefont{Z.}~\bibnamefont{Wang}},
  \bibinfo{author}{\bibfnamefont{W.-J.} \bibnamefont{Hu}}, \bibnamefont{and}
  \bibinfo{author}{\bibfnamefont{A.~H.} \bibnamefont{Nevidomskyy}},
  \bibinfo{journal}{Phys. Rev. Lett.} \textbf{\bibinfo{volume}{116}},
  \bibinfo{pages}{247203} (\bibinfo{year}{2016}),
  \urlprefix\url{https://link.aps.org/doi/10.1103/PhysRevLett.116.247203}.

\bibitem[{\citenamefont{Gong et~al.}(2017)\citenamefont{Gong, Zhu, Sheng, and
  Yang}}]{gong2017}
\bibinfo{author}{\bibfnamefont{S.-S.} \bibnamefont{Gong}},
  \bibinfo{author}{\bibfnamefont{W.}~\bibnamefont{Zhu}},
  \bibinfo{author}{\bibfnamefont{D.~N.} \bibnamefont{Sheng}}, \bibnamefont{and}
  \bibinfo{author}{\bibfnamefont{K.}~\bibnamefont{Yang}},
  \bibinfo{journal}{Phys. Rev. B} \textbf{\bibinfo{volume}{95}},
  \bibinfo{pages}{205132} (\bibinfo{year}{2017}),
  \urlprefix\url{https://link.aps.org/doi/10.1103/PhysRevB.95.205132}.

\bibitem[{\citenamefont{Lai et~al.}(2017)\citenamefont{Lai, Hu, Nica, Yu, and
  Si}}]{lai2017}
\bibinfo{author}{\bibfnamefont{H.-H.} \bibnamefont{Lai}},
  \bibinfo{author}{\bibfnamefont{W.-J.} \bibnamefont{Hu}},
  \bibinfo{author}{\bibfnamefont{E.~M.} \bibnamefont{Nica}},
  \bibinfo{author}{\bibfnamefont{R.}~\bibnamefont{Yu}}, \bibnamefont{and}
  \bibinfo{author}{\bibfnamefont{Q.}~\bibnamefont{Si}}, \bibinfo{journal}{Phys.
  Rev. Lett.} \textbf{\bibinfo{volume}{118}}, \bibinfo{pages}{176401}
  (\bibinfo{year}{2017}),
  \urlprefix\url{https://link.aps.org/doi/10.1103/PhysRevLett.118.176401}.

\bibitem[{\citenamefont{Haldane}(1983{\natexlab{a}})}]{Haldane1983_2}
\bibinfo{author}{\bibfnamefont{F.~D.~M.} \bibnamefont{Haldane}},
  \bibinfo{journal}{Phys. Rev. Lett.} \textbf{\bibinfo{volume}{50}},
  \bibinfo{pages}{1153} (\bibinfo{year}{1983}{\natexlab{a}}),
  \urlprefix\url{http://link.aps.org/doi/10.1103/PhysRevLett.50.1153}.

\bibitem[{\citenamefont{Affleck et~al.}(1987)\citenamefont{Affleck, Kennedy,
  Lieb, and Tasaki}}]{aklt1987}
\bibinfo{author}{\bibfnamefont{I.}~\bibnamefont{Affleck}},
  \bibinfo{author}{\bibfnamefont{T.}~\bibnamefont{Kennedy}},
  \bibinfo{author}{\bibfnamefont{E.~H.} \bibnamefont{Lieb}}, \bibnamefont{and}
  \bibinfo{author}{\bibfnamefont{H.}~\bibnamefont{Tasaki}},
  \bibinfo{journal}{Phys. Rev. Lett.} \textbf{\bibinfo{volume}{59}},
  \bibinfo{pages}{799} (\bibinfo{year}{1987}),
  \urlprefix\url{http://link.aps.org/doi/10.1103/PhysRevLett.59.799}.

\bibitem[{\citenamefont{Katsumata et~al.}(1989)\citenamefont{Katsumata, Hori,
  Takeuchi, Date, Yamagishi, and Renard}}]{katsumata1989}
\bibinfo{author}{\bibfnamefont{K.}~\bibnamefont{Katsumata}},
  \bibinfo{author}{\bibfnamefont{H.}~\bibnamefont{Hori}},
  \bibinfo{author}{\bibfnamefont{T.}~\bibnamefont{Takeuchi}},
  \bibinfo{author}{\bibfnamefont{M.}~\bibnamefont{Date}},
  \bibinfo{author}{\bibfnamefont{A.}~\bibnamefont{Yamagishi}},
  \bibnamefont{and} \bibinfo{author}{\bibfnamefont{J.~P.}
  \bibnamefont{Renard}}, \bibinfo{journal}{Phys. Rev. Lett.}
  \textbf{\bibinfo{volume}{63}}, \bibinfo{pages}{86} (\bibinfo{year}{1989}),
  \urlprefix\url{http://link.aps.org/doi/10.1103/PhysRevLett.63.86}.

\bibitem[{\citenamefont{Hagiwara et~al.}(1990)\citenamefont{Hagiwara,
  Katsumata, Affleck, Halperin, and Renard}}]{hagiwara1990}
\bibinfo{author}{\bibfnamefont{M.}~\bibnamefont{Hagiwara}},
  \bibinfo{author}{\bibfnamefont{K.}~\bibnamefont{Katsumata}},
  \bibinfo{author}{\bibfnamefont{I.}~\bibnamefont{Affleck}},
  \bibinfo{author}{\bibfnamefont{B.~I.} \bibnamefont{Halperin}},
  \bibnamefont{and} \bibinfo{author}{\bibfnamefont{J.~P.}
  \bibnamefont{Renard}}, \bibinfo{journal}{Phys. Rev. Lett.}
  \textbf{\bibinfo{volume}{65}}, \bibinfo{pages}{3181} (\bibinfo{year}{1990}),
  \urlprefix\url{http://link.aps.org/doi/10.1103/PhysRevLett.65.3181}.

\bibitem[{\citenamefont{White and Huse}(1993)}]{white1993}
\bibinfo{author}{\bibfnamefont{S.~R.} \bibnamefont{White}} \bibnamefont{and}
  \bibinfo{author}{\bibfnamefont{D.~A.} \bibnamefont{Huse}},
  \bibinfo{journal}{Phys. Rev. B} \textbf{\bibinfo{volume}{48}},
  \bibinfo{pages}{3844} (\bibinfo{year}{1993}),
  \urlprefix\url{http://link.aps.org/doi/10.1103/PhysRevB.48.3844}.

\bibitem[{\citenamefont{Schollw\"ock et~al.}(1996)\citenamefont{Schollw\"ock,
  Jolic\oe{}ur, and Garel}}]{schollwock1996}
\bibinfo{author}{\bibfnamefont{U.}~\bibnamefont{Schollw\"ock}},
  \bibinfo{author}{\bibfnamefont{T.}~\bibnamefont{Jolic\oe{}ur}},
  \bibnamefont{and} \bibinfo{author}{\bibfnamefont{T.}~\bibnamefont{Garel}},
  \bibinfo{journal}{Phys. Rev. B} \textbf{\bibinfo{volume}{53}},
  \bibinfo{pages}{3304} (\bibinfo{year}{1996}),
  \urlprefix\url{http://link.aps.org/doi/10.1103/PhysRevB.53.3304}.

\bibitem[{\citenamefont{Shelton et~al.}(1996)\citenamefont{Shelton, Nersesyan,
  and Tsvelik}}]{shelton1996}
\bibinfo{author}{\bibfnamefont{D.~G.} \bibnamefont{Shelton}},
  \bibinfo{author}{\bibfnamefont{A.~A.} \bibnamefont{Nersesyan}},
  \bibnamefont{and} \bibinfo{author}{\bibfnamefont{A.~M.}
  \bibnamefont{Tsvelik}}, \bibinfo{journal}{Phys. Rev. B}
  \textbf{\bibinfo{volume}{53}}, \bibinfo{pages}{8521} (\bibinfo{year}{1996}),
  \urlprefix\url{http://link.aps.org/doi/10.1103/PhysRevB.53.8521}.

\bibitem[{\citenamefont{L\"auchli
  et~al.}(2006{\natexlab{a}})\citenamefont{L\"auchli, Schmid, and
  Trebst}}]{lauchli2006}
\bibinfo{author}{\bibfnamefont{A.}~\bibnamefont{L\"auchli}},
  \bibinfo{author}{\bibfnamefont{G.}~\bibnamefont{Schmid}}, \bibnamefont{and}
  \bibinfo{author}{\bibfnamefont{S.}~\bibnamefont{Trebst}},
  \bibinfo{journal}{Phys. Rev. B} \textbf{\bibinfo{volume}{74}},
  \bibinfo{pages}{144426} (\bibinfo{year}{2006}{\natexlab{a}}),
  \urlprefix\url{http://link.aps.org/doi/10.1103/PhysRevB.74.144426}.

\bibitem[{\citenamefont{Corboz et~al.}(2007)\citenamefont{Corboz, L\"auchli,
  Totsuka, and Tsunetsugu}}]{corboz2007}
\bibinfo{author}{\bibfnamefont{P.}~\bibnamefont{Corboz}},
  \bibinfo{author}{\bibfnamefont{A.~M.} \bibnamefont{L\"auchli}},
  \bibinfo{author}{\bibfnamefont{K.}~\bibnamefont{Totsuka}}, \bibnamefont{and}
  \bibinfo{author}{\bibfnamefont{H.}~\bibnamefont{Tsunetsugu}},
  \bibinfo{journal}{Phys. Rev. B} \textbf{\bibinfo{volume}{76}},
  \bibinfo{pages}{220404} (\bibinfo{year}{2007}),
  \urlprefix\url{http://link.aps.org/doi/10.1103/PhysRevB.76.220404}.

\bibitem[{\citenamefont{Niesen and Corboz}(2017{\natexlab{a}})}]{corboz2017}
\bibinfo{author}{\bibfnamefont{I.}~\bibnamefont{Niesen}} \bibnamefont{and}
  \bibinfo{author}{\bibfnamefont{P.}~\bibnamefont{Corboz}},
  \bibinfo{journal}{Phys. Rev. B} \textbf{\bibinfo{volume}{95}},
  \bibinfo{pages}{180404} (\bibinfo{year}{2017}{\natexlab{a}}),
  \urlprefix\url{https://link.aps.org/doi/10.1103/PhysRevB.95.180404}.

\bibitem[{\citenamefont{Haldane}(1983{\natexlab{b}})}]{Haldane1983_1}
\bibinfo{author}{\bibfnamefont{F.}~\bibnamefont{Haldane}},
  \bibinfo{journal}{Physics Letters A} \textbf{\bibinfo{volume}{93}},
  \bibinfo{pages}{464 } (\bibinfo{year}{1983}{\natexlab{b}}),
  \urlprefix\url{http://www.sciencedirect.com/science/article/pii/037596018390631X}.

\bibitem[{\citenamefont{Lieb et~al.}(1961)\citenamefont{Lieb, Schultz, and
  Mattis}}]{lieb1961}
\bibinfo{author}{\bibfnamefont{E.}~\bibnamefont{Lieb}},
  \bibinfo{author}{\bibfnamefont{T.}~\bibnamefont{Schultz}}, \bibnamefont{and}
  \bibinfo{author}{\bibfnamefont{D.}~\bibnamefont{Mattis}},
  \bibinfo{journal}{Annals of Physics} \textbf{\bibinfo{volume}{16}},
  \bibinfo{pages}{407} (\bibinfo{year}{1961}),
  \urlprefix\url{https://doi.org/10.1016/0003-4916(61)90115-4}.

\bibitem[{\citenamefont{Hastings}(2004)}]{Hastings2004}
\bibinfo{author}{\bibfnamefont{M.~B.} \bibnamefont{Hastings}},
  \bibinfo{journal}{Phys. Rev. B} \textbf{\bibinfo{volume}{69}},
  \bibinfo{pages}{104431} (\bibinfo{year}{2004}),
  \urlprefix\url{http://link.aps.org/doi/10.1103/PhysRevB.69.104431}.

\bibitem[{\citenamefont{Changlani and L\"auchli}(2015)}]{changlani}
\bibinfo{author}{\bibfnamefont{H.~J.} \bibnamefont{Changlani}}
  \bibnamefont{and} \bibinfo{author}{\bibfnamefont{A.~M.}
  \bibnamefont{L\"auchli}}, \bibinfo{journal}{Phys. Rev. B}
  \textbf{\bibinfo{volume}{91}}, \bibinfo{pages}{100407}
  (\bibinfo{year}{2015}),
  \urlprefix\url{https://link.aps.org/doi/10.1103/PhysRevB.91.100407}.

\bibitem[{\citenamefont{L\"auchli
  et~al.}(2006{\natexlab{b}})\citenamefont{L\"auchli, Mila, and
  Penc}}]{lauchli2006_2}
\bibinfo{author}{\bibfnamefont{A.}~\bibnamefont{L\"auchli}},
  \bibinfo{author}{\bibfnamefont{F.}~\bibnamefont{Mila}}, \bibnamefont{and}
  \bibinfo{author}{\bibfnamefont{K.}~\bibnamefont{Penc}},
  \bibinfo{journal}{Phys. Rev. Lett.} \textbf{\bibinfo{volume}{97}},
  \bibinfo{pages}{087205} (\bibinfo{year}{2006}{\natexlab{b}}),
  \urlprefix\url{https://link.aps.org/doi/10.1103/PhysRevLett.97.087205}.

\bibitem[{\citenamefont{T\'oth et~al.}(2010)\citenamefont{T\'oth, L\"auchli,
  Mila, and Penc}}]{toth2010}
\bibinfo{author}{\bibfnamefont{T.~A.} \bibnamefont{T\'oth}},
  \bibinfo{author}{\bibfnamefont{A.~M.} \bibnamefont{L\"auchli}},
  \bibinfo{author}{\bibfnamefont{F.}~\bibnamefont{Mila}}, \bibnamefont{and}
  \bibinfo{author}{\bibfnamefont{K.}~\bibnamefont{Penc}},
  \bibinfo{journal}{Phys. Rev. Lett.} \textbf{\bibinfo{volume}{105}},
  \bibinfo{pages}{265301} (\bibinfo{year}{2010}),
  \urlprefix\url{http://link.aps.org/doi/10.1103/PhysRevLett.105.265301}.

\bibitem[{\citenamefont{T\'oth et~al.}(2012)\citenamefont{T\'oth, L\"auchli,
  Mila, and Penc}}]{toth2012}
\bibinfo{author}{\bibfnamefont{T.~A.} \bibnamefont{T\'oth}},
  \bibinfo{author}{\bibfnamefont{A.~M.} \bibnamefont{L\"auchli}},
  \bibinfo{author}{\bibfnamefont{F.}~\bibnamefont{Mila}}, \bibnamefont{and}
  \bibinfo{author}{\bibfnamefont{K.}~\bibnamefont{Penc}},
  \bibinfo{journal}{Phys. Rev. B} \textbf{\bibinfo{volume}{85}},
  \bibinfo{pages}{140403} (\bibinfo{year}{2012}),
  \urlprefix\url{http://link.aps.org/doi/10.1103/PhysRevB.85.140403}.

\bibitem[{\citenamefont{Bauer et~al.}(2012)\citenamefont{Bauer, Corboz,
  L\"auchli, Messio, Penc, Troyer, and Mila}}]{bauer2012}
\bibinfo{author}{\bibfnamefont{B.}~\bibnamefont{Bauer}},
  \bibinfo{author}{\bibfnamefont{P.}~\bibnamefont{Corboz}},
  \bibinfo{author}{\bibfnamefont{A.~M.} \bibnamefont{L\"auchli}},
  \bibinfo{author}{\bibfnamefont{L.}~\bibnamefont{Messio}},
  \bibinfo{author}{\bibfnamefont{K.}~\bibnamefont{Penc}},
  \bibinfo{author}{\bibfnamefont{M.}~\bibnamefont{Troyer}}, \bibnamefont{and}
  \bibinfo{author}{\bibfnamefont{F.}~\bibnamefont{Mila}},
  \bibinfo{journal}{Phys. Rev. B} \textbf{\bibinfo{volume}{85}},
  \bibinfo{pages}{125116} (\bibinfo{year}{2012}),
  \urlprefix\url{http://link.aps.org/doi/10.1103/PhysRevB.85.125116}.

\bibitem[{\citenamefont{Zhao et~al.}(2012)\citenamefont{Zhao, Xu, Chen, Wei,
  Qin, Zhang, and Xiang}}]{zhao2012}
\bibinfo{author}{\bibfnamefont{H.~H.} \bibnamefont{Zhao}},
  \bibinfo{author}{\bibfnamefont{C.}~\bibnamefont{Xu}},
  \bibinfo{author}{\bibfnamefont{Q.~N.} \bibnamefont{Chen}},
  \bibinfo{author}{\bibfnamefont{Z.~C.} \bibnamefont{Wei}},
  \bibinfo{author}{\bibfnamefont{M.~P.} \bibnamefont{Qin}},
  \bibinfo{author}{\bibfnamefont{G.~M.} \bibnamefont{Zhang}}, \bibnamefont{and}
  \bibinfo{author}{\bibfnamefont{T.}~\bibnamefont{Xiang}},
  \bibinfo{journal}{Phys. Rev. B} \textbf{\bibinfo{volume}{85}},
  \bibinfo{pages}{134416} (\bibinfo{year}{2012}),
  \urlprefix\url{https://link.aps.org/doi/10.1103/PhysRevB.85.134416}.

\bibitem[{\citenamefont{Corboz et~al.}(2012)\citenamefont{Corboz, Penc, Mila,
  and L\"auchli}}]{corboz2012}
\bibinfo{author}{\bibfnamefont{P.}~\bibnamefont{Corboz}},
  \bibinfo{author}{\bibfnamefont{K.}~\bibnamefont{Penc}},
  \bibinfo{author}{\bibfnamefont{F.}~\bibnamefont{Mila}}, \bibnamefont{and}
  \bibinfo{author}{\bibfnamefont{A.~M.} \bibnamefont{L\"auchli}},
  \bibinfo{journal}{Phys. Rev. B} \textbf{\bibinfo{volume}{86}},
  \bibinfo{pages}{041106} (\bibinfo{year}{2012}),
  \urlprefix\url{http://link.aps.org/doi/10.1103/PhysRevB.86.041106}.

\bibitem[{\citenamefont{Corboz et~al.}(2013)\citenamefont{Corboz, Lajk\'o,
  Penc, Mila, and L\"auchli}}]{corboz2013}
\bibinfo{author}{\bibfnamefont{P.}~\bibnamefont{Corboz}},
  \bibinfo{author}{\bibfnamefont{M.}~\bibnamefont{Lajk\'o}},
  \bibinfo{author}{\bibfnamefont{K.}~\bibnamefont{Penc}},
  \bibinfo{author}{\bibfnamefont{F.}~\bibnamefont{Mila}}, \bibnamefont{and}
  \bibinfo{author}{\bibfnamefont{A.~M.} \bibnamefont{L\"auchli}},
  \bibinfo{journal}{Phys. Rev. B} \textbf{\bibinfo{volume}{87}},
  \bibinfo{pages}{195113} (\bibinfo{year}{2013}),
  \urlprefix\url{http://link.aps.org/doi/10.1103/PhysRevB.87.195113}.

\bibitem[{\citenamefont{Niesen and
  Corboz}(2017{\natexlab{b}})}]{niesen2017tensor}
\bibinfo{author}{\bibfnamefont{I.}~\bibnamefont{Niesen}} \bibnamefont{and}
  \bibinfo{author}{\bibfnamefont{P.}~\bibnamefont{Corboz}},
  \bibinfo{journal}{SciPost Phys.} \textbf{\bibinfo{volume}{3}},
  \bibinfo{pages}{030} (\bibinfo{year}{2017}{\natexlab{b}}),
  \urlprefix\url{https://scipost.org/10.21468/SciPostPhys.3.4.030}.

\bibitem[{\citenamefont{Niesen and Corboz}(2018)}]{corboz2018}
\bibinfo{author}{\bibfnamefont{I.}~\bibnamefont{Niesen}} \bibnamefont{and}
  \bibinfo{author}{\bibfnamefont{P.}~\bibnamefont{Corboz}},
  \bibinfo{journal}{Phys. Rev. B} \textbf{\bibinfo{volume}{97}},
  \bibinfo{pages}{245146} (\bibinfo{year}{2018}),
  \urlprefix\url{https://link.aps.org/doi/10.1103/PhysRevB.97.245146}.

\bibitem[{\citenamefont{Wang et~al.}(2019)\citenamefont{Wang, Hu, Yu, and
  Si}}]{b1gb2g}
\bibinfo{author}{\bibfnamefont{Y.}~\bibnamefont{Wang}},
  \bibinfo{author}{\bibfnamefont{W.}~\bibnamefont{Hu}},
  \bibinfo{author}{\bibfnamefont{R.}~\bibnamefont{Yu}}, \bibnamefont{and}
  \bibinfo{author}{\bibfnamefont{Q.}~\bibnamefont{Si}}, \bibinfo{journal}{Phys.
  Rev. B} \textbf{\bibinfo{volume}{100}}, \bibinfo{pages}{100502}
  (\bibinfo{year}{2019}),
  \urlprefix\url{https://link.aps.org/doi/10.1103/PhysRevB.100.100502}.

\bibitem[{\citenamefont{White}(1992)}]{white1992}
\bibinfo{author}{\bibfnamefont{S.~R.} \bibnamefont{White}},
  \bibinfo{journal}{Phys. Rev. Lett.} \textbf{\bibinfo{volume}{69}},
  \bibinfo{pages}{2863} (\bibinfo{year}{1992}),
  \urlprefix\url{http://link.aps.org/doi/10.1103/PhysRevLett.69.2863}.

\bibitem[{\citenamefont{McCulloch and Gul{\'a}csi}(2002)}]{mcculloch2002}
\bibinfo{author}{\bibfnamefont{I.}~\bibnamefont{McCulloch}} \bibnamefont{and}
  \bibinfo{author}{\bibfnamefont{M.}~\bibnamefont{Gul{\'a}csi}},
  \bibinfo{journal}{Europhysics Letters} \textbf{\bibinfo{volume}{57}},
  \bibinfo{pages}{852} (\bibinfo{year}{2002}),
  \urlprefix\url{http://iopscience.iop.org/0295-5075/57/6/852}.

\bibitem[{\citenamefont{Blume and Hsieh}(1969)}]{blume1969}
\bibinfo{author}{\bibfnamefont{M.}~\bibnamefont{Blume}} \bibnamefont{and}
  \bibinfo{author}{\bibfnamefont{Y.}~\bibnamefont{Hsieh}},
  \bibinfo{journal}{Journal of Applied Physics} \textbf{\bibinfo{volume}{40}},
  \bibinfo{pages}{1249} (\bibinfo{year}{1969}),
  \urlprefix\url{https://doi.org/10.1063/1.1657616}.

\bibitem[{\citenamefont{Chandra et~al.}(1990)\citenamefont{Chandra, Coleman,
  and Larkin}}]{Chandra1990}
\bibinfo{author}{\bibfnamefont{P.}~\bibnamefont{Chandra}},
  \bibinfo{author}{\bibfnamefont{P.}~\bibnamefont{Coleman}}, \bibnamefont{and}
  \bibinfo{author}{\bibfnamefont{A.~I.} \bibnamefont{Larkin}},
  \bibinfo{journal}{Phys. Rev. Lett.} \textbf{\bibinfo{volume}{64}},
  \bibinfo{pages}{88} (\bibinfo{year}{1990}),
  \urlprefix\url{https://link.aps.org/doi/10.1103/PhysRevLett.64.88}.

\end{thebibliography}

\end{document}